\documentclass[a4paper]{article}

\usepackage[top=25truemm,bottom=25truemm,left=25truemm,right=25truemm]{geometry}

\usepackage{fancyhdr}
\usepackage{fancybox}
\usepackage{ascmac} 
\usepackage{amsmath}
\usepackage{mathtools}
\usepackage{amsfonts}
\usepackage{amssymb}%
\usepackage{mathrsfs}%
\usepackage{amsthm}%
\usepackage{bm} %
\usepackage{enumerate}
\usepackage{lastpage}
\usepackage{graphicx,color}
\usepackage{version}
\usepackage{comment}
\usepackage{cancel}
\usepackage{url}%
\usepackage{cite}

\allowdisplaybreaks

\newcommand{\drm}{d} 
 
\newcommand{\irm}{i}

\newcommand{\Exp}[1]{\mathrm{e}^{#1}}

\newcommand{\Trm}{\mathrm{T}}

\newcommand{\Tr}[1]{\mathrm{Tr}{#1}}

\newcommand{\Slash}[1]{{\ooalign{\hfil/\hfil\crcr$#1$}}}

\newcommand{\bra}[1]{\langle #1 |}
\newcommand{\ket}[1]{| #1 \rangle}

\newcommand{\lpartial}{\overset{\leftarrow}{\partial}}

\newcommand{\cl}{\mathrm{c.l.}}
\newcommand{\beq}{\begin{eqnarray}}
\newcommand{\eeq}{\end{eqnarray}}
\newcommand{\nn}{\nonumber}

\def\zh{\hat{z}}

\def\xh{\hat{x}}

\begin{document}
\begin{flushright}
\end{flushright}
\vspace*{5mm}
\begin{center}
{\large \bf
Matching between
Collinear Twist-3 and TMD Fragmentation Function\\[10pt]  
Contributions to
Polarized
Hyperon Production 
in SIDIS
}
\vspace{1.5cm}\\
{\sc Riku Ikarashi$^1$, Yuji Koike$^2$, and Shinsuke Yoshida$^{3,4}$}
\\[0.7cm]
\vspace*{0.1cm}
{\it $^1$Graduate School of Science and Technology, Niigata University,
Ikarashi 2-no-cho, Niigata 950-2181, Japan}

\vspace{0.2cm}

{\it $^2$Department of Physics, Niigata University, Ikarashi 2-no-cho, Niigata 950-2181, Japan}

\vspace{0.2cm}

{\it $^3$State Key Laboratory of Nuclear Physics and
Technology, Institute of Quantum Matter, 
South China Normal University, Guangzhou 510006, China}

\vspace{0.2cm}

{\it $^4$Guangdong Basic Research Center of Excellence for
Structure and Fundamental Interactions of Matter,
Guangdong Provincial Key Laboratory of Nuclear Science,
Guangzhou 510006, China}
\\[3cm]

{\large \bf Abstract} 
\end{center}
We investigate the consistency between the collinear twist-3 factorization and the 
transverse-momentum-dependent (TMD) factorization
for the transversely polarized hyperon production
in semi-inclusive deep inelastic scattering, $ep\to e\Lambda^\uparrow X$. 
In particular, we focus on the contributions from the
twist-3 fragmentation functions (FFs) and the TMD polarizing FF
in the region of the hyperon's intermediate transverse momentum $P_T$,
$\Lambda_{\rm QCD}\ll P_T\ll Q$, where both frameworks are valid.  
In this region the polarizing FF can be expressed in terms of the twist-3 FFs including the
purely gluonic ones, 
and the resulting TMD factorization formula for $ep\to e\Lambda^\uparrow X$
agrees with the small-$P_T$ limit of the corresponding twist-3 cross section.  
This matching of the two calculations indicates that the two frameworks describe 
the same effect in QCD
and provide complementary frameworks for the process in different kinematic regions.  

\newpage

\section{Introduction}

Understanding the mechanism for the
transverse polarization of hyperons produced in unpolarized 
collisions\footnote{In this paper we collectively denote spin-1/2 hyperon as $\Lambda$.}, 
such as $pp\to \Lambda^\uparrow X$, $ep\to (e)\Lambda^\uparrow X$ and $e^+e^-\to
\Lambda^\uparrow X$ etc., 
has been one of the central issues in high-energy QCD.  
This hyperon polarization, as well as $p^\uparrow p\to h X$, $ep^\uparrow \to (e)h X$ 
($h=\pi,\ \gamma,\ {\rm jet}$
etc.),
are the typical examples of the transverse single spin asymmetry (SSA), 
and have been studied at length in the literature.  
Collinear twist-2 factorization in perturbative QCD, which has been successful
in describing many unpolarized collision processes, 
cannot accommodate large SSAs\cite{Kane:1978nd}.  
It has been understood by now that effects of particular multi-parton 
correlations and/or intrinsic transverse momentum of partons
either in the initial hadrons or in the fragmenting process
can cause large SSAs.
Systematic methods of handling those effects have been formulated in the framework of
the collinear twist-3 
factorization\cite{Efremov:1981sh,Qiu:1991wg,Qiu:1998ia,Eguchi:2006qz,Eguchi:2006mc,
Kouvaris:2006zy,Koike:2009ge,Koike:2009yb,Beppu:2010qn,Metz:2012ct,Kanazawa:2013uia,
Koike:2017fxr,Koike:2021awj} 
and the transverse momentum dependent (TMD) factorization\cite{Ji:2004xq,Ji:2004wu}.  
Collinear twist-3 factorization is designed to describe SSAs in the 
region of the large transverse momentum of the final hadron $\Lambda_{\rm QCD}\ll P_T \lesssim Q$
where $Q$ is a typical hard scale in the process,
while the TMD factorization describes those in the small-$P_T$ region $\Lambda_{\rm QCD}\lesssim P_T \ll Q$.  
Relation between these two frameworks has also been studied for some processes, and it's been shown that
they match in the region of the intermediate transverse momentum ($\Lambda_{\rm QCD}\ll P_T \ll Q$)
for $ep^\uparrow\to e\pi X$ and Drell-Yan processes etc\cite{Ji:2006ub,Ji:2006br,Ji:2006vf,Koike:2007dg,Zhou:2008,Yuan:2009dw}.
Although the twist-3
calculation so far has been mostly limited to the leading order (LO) with respect to the
QCD coupling due to the technical complexities, 
there have been some studies to include 
next-to-leading order (NLO)
corrections 
for $ep^\uparrow\to e\pi X$\cite{Kang:2012ns,
Dai:2014ala,Yoshida:2016tfh,Chen:2017lvx,Benic:2019zvg},
Drell-Yan process\cite{Vogelsang:2009pj,Chen:2016dnp},
and $e^+e^-\to \Lambda^\uparrow X$\cite{Gamberg:2018fwy} etc.  
More recently, a complete NLO twist-3 distribution function (DF) 
contribution to the fully differential
cross section for
$ep^\uparrow\to hX$ ($h=\pi$, jet) has been derived and its factorization property
has been proven\cite{Rein:2025pwu,Rein:2025qhe}.  

In this paper we investigate the consistency between the collinear twist-3
factorization and the TMD factorization at LO
in the intermediate transverse momentum region
for the 
hyperon polarization in semi-inclusive deep inelastic scattering (SIDIS) $ep\to e\Lambda^\uparrow X$.  
In the collinear twist-3 factorization, two effects participate in this process:  One is
(i) the twist-3 unpolarized DF in the initial proton 
convoluted with the twist-2 transversity fragmentation function (FF) for the final hyperon 
and the corresponding partonic hard cross section, and the other is 
(ii) the twist-3 FFs for the final hyperon convoluted with
the unpolarized DF in the proton and the partonic hard cross section.  
In (ii), two contributions exist; (a) the twist-3 quark FFs and (b) the 
twist-3 gluon FFs.  
The complete LO twist-3 cross section formula
for $ep\to e\Lambda^\uparrow X$ has been derived 
for
(i) in \cite{Zhou:2008,Koike:2022ddx}, for  (ii)(a)
 in \cite{Koike:2022ddx} and for (ii)(b) in \cite{Ikarashi:2022yzg}.  
In the TMD factorization, corresponding 
to the above (i) and (ii), there are two contributions: 
(I) the Boer-Mulders function in the initial proton combined with
the transversity FF for the hyperon\cite{Boer:1997nt} 
and (II) the polarizing FF combined with
the unpolarized DF in the proton contribute to 
the hyperon polarization\cite{Mulders:1995dh}.  
So far, it's been shown that (i) and (I) match at intermediate transverse 
momentum $\Lambda_{\rm QCD}\ll P_T \ll Q$\cite{Zhou:2008},
while no such relation has been established between (ii) and (II).

The purpose of this paper is to investigate the relation between (ii) and (II)
for $ep\to e\Lambda^\uparrow X$.
So far such study on the matching between the two frameworks
has been mostly for the distribution side, 
and the study on the FF contribution to SSA has been performed only for the process 
$ep^\uparrow\to e\pi X$ in \cite{Yuan:2009dw}, which shows
the twist-3 FF contribution matches with that from the Collins 
function\cite{Collins:1992kk}.  
This paper extends the study to $ep\to e\Lambda^\uparrow X$.  
For (ii) and (II), extra complication occurs
compared to $ep^\uparrow\to e\pi X$:
Owing to the chiral-even nature of the relevant twist-3 FFs and the polarizing FF,
a purely gluonic channel (above (ii)(b)) enters the calculation, 
for which the complete twist-3 cross 
section was derived only recently\cite{Ikarashi:2022yzg}.  
Here we will show that the TMD polarizing FF contribution can be expressed 
in terms of the collinear twist-3 quark and gluon FFs in the region of 
large intrinsic transverse momentum,
and the resulting TMD factorization formula agrees with 
that obtained from the small-$P_T$ limit
of the collinear twist-3 cross section.  

This paper is organized as follows:
In section 2, we summarize the kinematics for $ep\to e\Lambda^\uparrow X$.    
In section 3, we summarize the collinear twist-3 FFs and the TMD polarizing FF relevant to the present study.  
In section 4, we recall the structure of the twist-3 FF contribution for $ep\to e\Lambda^\uparrow X$
from \cite{Koike:2022ddx,Ikarashi:2022yzg}
and derive its small-$P_T$ limit.
In section 5, we present a perturbative calculation of the TMD polarizing FF in the 
region of the large intrinsic transverse momentum,
and show that the two frameworks indeed match at $\Lambda_{\rm QCD}\ll P_T \ll Q$.  
Section 6 is devoted to a brief summary.

\section{Kinematics for $ep\to e\Lambda^\uparrow X$}

Here we briefly summarize the kinematics for 
the process following \cite{Koike:2022ddx,Ikarashi:2022yzg},
\beq
e(\ell) + p (p) \to e (\ell') + \Lambda^{\uparrow}(P_h, S_\perp) + X,  
\eeq
where $\ell$, $\ell'$, $p$ and $P_h$ are the momenta of each particle and
$S_\perp$ is the transverse spin vector for $\Lambda$.  With the
virtual photon's momentum $q=\ell-\ell'$, 
we introduce the five Lorentz invariants as
\beq
&&S_{ep} \equiv (p+\ell)^2 \simeq 2 p \cdot \ell,\qquad
Q^2 \equiv -q^2,
\nonumber\\
&&x_{bj} \equiv \frac{Q^2}{2 p \cdot q} , \qquad 
z_f \equiv \frac{p \cdot P_h}{p \cdot q},\qquad
q_T \equiv \sqrt{-q_t^2},
\eeq
where
\beq
q_t^\mu \equiv q^\mu - \frac{P_h \cdot q}{p \cdot P_h}p^\mu
-\frac{p \cdot q}{p \cdot P_h} P_h^\mu
\eeq
is a space-like momentum satisfying $q_t \cdot p = q_t \cdot P_h = 0$.
\begin{figure}[h]
 \begin{center}
\includegraphics[scale=0.9]{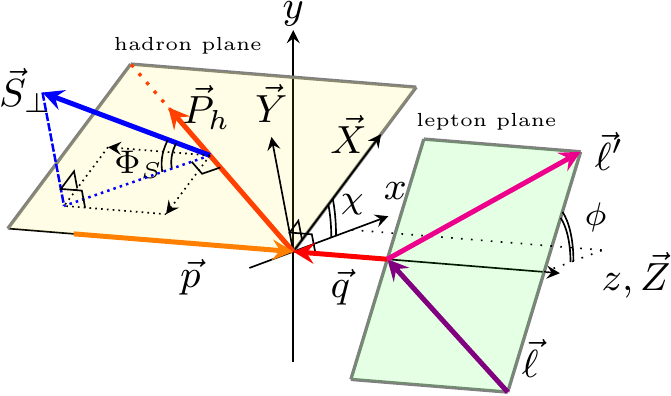}
\caption{Hadron frame}
\label{hadronframe}
\end{center}
\end{figure}
As in \cite{Koike:2022ddx,Ikarashi:2022yzg}, we work in the hadron 
frame\cite{Meng:1991da} (See Fig. \ref{hadronframe}),
where $p^\mu$ and $q^\mu$ are collinear and take the following form: 
\begin{align}
&p^\mu = \frac{Q}{2 x_{bj}}(1,0,0,1),\\
&q^\mu = (0,0,0,-Q). 
\end{align}
Defining the azimuthal angles for the hadron plane and the lepton plane
as $\chi$ and $\phi$, respectively, as shown in Fig. \ref{hadronframe}, 
$P_h^\mu$ and $\ell^\mu$ can be written as 
\beq
&& P_h^\mu = \frac{z_f Q}{2} \left(
1 + \frac{q_T^2}{Q^2}, \frac{2 q_T}{Q} \cos{\chi},
\frac{2 q_T}{Q} \sin{\chi}, -1 + \frac{q_T^2}{Q^2}
\right),
\label{Ph}\\
&&\ell^\mu = \frac{Q}{2}(\cosh{\psi}, \sinh{\psi}\cos{\phi},
\sinh{\psi}\sin{\phi},-1 ),
\label{lepmom}
\eeq
where $\cosh{\psi} \equiv \frac{2x_{bj} S_{ep}}{Q^2} -1$.  
With this parameterization, the magnitude of
the transverse momentum of the hyperon $P_{hT}\equiv P_{T}$
is given by
\begin{equation}
P_{T}= z_f q_T. 
\end{equation}

For the calculation of the cross section, 
we introduce four axes by
\begin{align}
\label{txyz}
 & T^\mu \equiv \frac{1}{Q}(q^\mu + 2 x_{bj} p^\mu)=(1,0,0,0),{\nonumber}\\
&Z^\mu \equiv - \frac{q^\mu}{Q} = (0,0,0,1), \nonumber\\
 &X^\mu \equiv \frac{1}{q_T} \left[
\frac{P_h^\mu}{z_f} - q^\mu - \left(
1+ \frac{q_T^2}{Q^2} 
\right)x_{bj} p^\mu
\right] = (0, \cos{\chi}, \sin{\chi}, 0) , {\nonumber}\\
 & Y^\mu \equiv \epsilon^{\mu \nu \rho \sigma}Z_\nu T_\rho X_\sigma  
= (0, - \sin{\chi}, \cos{\chi},0),
\end{align}
where the actual form in the hadron frame is given after the last equality in each equation. 
The final hyperon resides in the $XZ$-plane and the transverse spin vector of the hyperon
can be written as
\begin{align}
\label{spinvector}
 S_\perp^\mu = \cos{\theta}\cos{\Phi_S}X^\mu + \sin{\Phi_S}Y^\mu - \sin{\theta}\cos{\Phi_S}Z^\mu, 
\end{align}
where $\theta$ is the polar angle of $\vec{P_h}$ as measured from the $Z$-axis
and $\Phi_S$ is the azimuthal angle of $\vec{S}_\perp$ around $\vec{P}_h$
as measured from the $XZ$-plane.   
From (\ref{Ph}), the polar angle $\theta$ is written as
\begin{align}
 \cos{\theta} &= \frac{P_{hz}}{|\vec P_h|}
= \frac{q_T^2 - Q^2}{q_T^2 + Q^2}, \\
 \sin{\theta}&=   \frac{P_{T}}{|\vec P_h |}
= \frac{2 q_T Q}{q_T^2 + Q^2}. 
\end{align}

With the kinematical variables defined above,
the differential cross section takes the following form:
\begin{align}
\label{Xsec}
 \frac{\drm^6 \sigma}{\drm x_{bj} \drm Q^2 \drm z_f \drm q_T^2 \drm \phi \drm \chi}
= \frac{\alpha_{em}^2}{128 \pi^4 x_{bj}^2 S_{ep}^2 Q^2}z_f
L^{\rho\sigma}(\ell , \ell')W_{\rho\sigma}(p,q,P_h) ,
\end{align}
where $\alpha_{em} = e^2/(4\pi)$ is the QED coupling constant, 
$ L^{\rho\sigma}=2(\ell^\rho \ell'^\sigma + \ell^\sigma \ell'^\rho)-Q^2 g^{\rho\sigma}$
is the leptonic tensor and $W_{\rho\sigma}$ is the hadronic tensor
which can be expanded in terms of
the six tensors\cite{Meng:1991da}
$\mathscr{V}_k^{\rho\sigma}$ ($k=1,\cdots,4, 8 ,9)$ defined by
\begin{align}
  &\mathscr{V}^{\mu \nu}_1 = X^\mu X^\nu + Y^\mu Y^\nu ,&
& \mathscr{V}^{\mu \nu}_2 = g^{\mu \nu}+ Z^\mu Z^\nu, &
& \mathscr{V}^{\mu \nu}_3 = T^\mu X^\nu + X^\mu T^\nu ,
{\nonumber}\\
  &\mathscr{V}^{\mu \nu}_4 = X^\mu X^\nu - Y^\mu Y^\nu ,&
& \mathscr{V}^{\mu \nu}_8 = T^\mu Y^\nu  + Y^\mu T^\nu,&
 &\mathscr{V}^{\mu \nu}_9 = X^\mu Y^\nu + Y^\mu X^\nu.  
\end{align}
Although there are two azimuthal angles, $\phi$ and $\chi$, the cross section depends on 
the relative angle $\varphi \equiv \phi - \chi$ only, and the $\varphi$-dependence of
$L^{\rho\sigma}W_{\rho\sigma}$ in (\ref{Xsec}) is carried by
the coefficients 
$\mathscr{A}_k(\varphi) \equiv L_{\mu \nu}\mathscr{V}^{\mu\nu}_k/Q^2$
which are given by
\begin{align}
& \mathscr{A}_1(\varphi) = 1 + \cosh^2{\psi}, &
&\mathscr{A}_2(\varphi) = -2, &
&\mathscr{A}_3(\varphi) = - \cos{\varphi}\sinh{2\psi},{\nonumber} \\
 &\mathscr{A}_4(\varphi) = \cos{2\varphi} \sinh^2{\psi} ,&
&\mathscr{A}_8(\varphi)  = - \sin{\varphi}\sinh{2\psi}, &
&\mathscr{A}_9(\varphi) = \sin{2\varphi} \sinh^2{\psi}.
\label{Aks}
\end{align}
The cross section (\ref{Xsec}) can be expressed in terms of 
$S_{ep}$, $Q^2$, $x_{bj}$, $z_f$, $q_T^2$, $\varphi$ and $\Phi_S$.

\section{Collinear twist-3 FFs and TMD FF for $ep\to e\Lambda^\uparrow X$}
Here we summarize the FFs contributing to $ep \to e\Lambda^\uparrow X$
in the collinear twist-3 factorization and the TMD factorization.  

\subsection{Collinear twist-3 FFs}
\subsubsection{Twist-3 quark FFs}
Three kinds of twist-3 quark FFs\cite{Kanazawa:2015ajw} contribute to 
$ep \to e\Lambda^\uparrow X$.

\vspace{3mm}

\noindent 
(1) Intrinsic FFs:

They are defined from the lightcone correlation function of the quark fields as
\begin{align}\label{quarkintFFdef}
 \Delta (z)&= \frac{1}{N}\sum_{X} \int \frac{\drm \lambda}{2\pi}
\Exp{-\irm \frac{\lambda}{z}}\bra{0}
\psi (0)\ket{\Lambda^\uparrow X}
\bra{\Lambda^\uparrow X} \bar \psi (\lambda w)\ket{0}
{\nonumber} \\
 & = \gamma_5 \Slash{S}_\perp \Slash{P}_h\frac{H_1(z)}{z}
+M_h \epsilon^{\alpha S_\perp w P_h} \gamma_\alpha\frac{D_T(z)}{z}
+M_h \gamma_5 \Slash{S}_\perp\frac{G_T(z)}{z}+\cdots,
\end{align}
where $z$ is the longitudinal momentum fraction transferred to the hyperon, and $N$ is the number of color SU(N), 
and $\ket{\Lambda^\uparrow }$ is the state vector for the hyperon
with mass $M_h$, momentum $P_h$ ($P_h^2= M_h^2$), and the transverse spin vector $S_\perp$ ($S_\perp^2=-1$).
In the twist-3 accuracy, $P_h$ can be treated as lightlike and 
$w$ is the another lightlike vector satisfying $P_h \cdot w = 1$, i.e.,
for $P_h = (|\vec P_h|, \vec P_h)$, 
$w= \frac{1}{2 |\vec P_h|^2}(|\vec P_h|, - \vec P_h)$.  
Here and below, we use the notation 
$\epsilon^{\alpha S_\perp w P_h}\equiv \epsilon^{\alpha\mu\nu\lambda}S_{\perp\mu}w_\nu P_{h\lambda}$ and
the gauge-link operators which restore the color-gauge invariance of FFs are implicit.  
$H_1 (z)$ is the twist-2 transversity fragmentation 
contributing to $ep \to e\Lambda^\uparrow X$ and $pp\to \Lambda^\uparrow X$
combined with the twist-3 unpolarized distribution in the nucleon
$E_F(x_1,x_2)$\cite{Koike:2022ddx,Kanazawa:2000hz,Kanazawa:2001a,Koike:2015zya}, 
which we shall not discuss here.  
$D_T (z)$ and $G_T(z)$ are the {\it intrinsic} twist-3 quark fragmentation function, 
in particular, 
$D_T(z)$ is T-odd and contributes to 
$ep \to e\Lambda^\uparrow X$.  
In (\ref{quarkintFFdef}) and below, the flavor indices for the quark FFs are suppressed for simplicity.  

\vspace{3mm}

\noindent
(2) Kinematical FFs:

Twist-3 kinematical quark fragmentation functions are defined from the 
correlation functions containing the transverse derivative of the fields:  
\begin{align}\label{quarkkinFFdef}
 \Delta^\alpha_{\partial}(z) &= \frac{1}{N} \sum_X 
\int\frac{\drm \lambda}{2\pi}
\Exp{-\irm  \frac{\lambda}{z}}\bra{0}
\psi (0) \ket{\Lambda^\uparrow X}
\bra{\Lambda^\uparrow X}\bar \psi (\lambda w)
\ket{0}\lpartial{}^\alpha {\nonumber}\\
 & =-i M_h \epsilon^{\alpha S_\perp w P_h} \Slash{P}_h
\frac{D_{1T}^{\perp (1)}(z)}{z}
+iM_h S_\perp^\alpha \gamma_5 \Slash{P}_h
\frac{G_{1T}^{\perp (1)}(z)}{z} +\cdots,
\end{align}
where
$\bar \psi (\lambda w) \lpartial{}^\alpha \equiv 
\lim_{\xi \to 0} (d/d\xi_\alpha) \bar \psi (\lambda w +\xi)$.  
$D_{1T}^{\perp (1)}(z)$ and $G_{1T}^{\perp (1)}(z)$
are called  {\it kinematical} fragmentation functions, 
with its superscript $(1)$ denoting the $k_T^2 / M_h^2$-moment of the
TMD fragmentation functions\,\cite{Mulders:1995dh}.  (See eq. (\ref{moment}) below.)
$G_{1T}^{\perp (1)}(z)$ is T-even, while
$D_{1T}^{\perp (1)}(z)$ is T-odd and contributes to $ep \to e\Lambda^\uparrow X$.  

\vspace{3mm}

\noindent
(3) Dynamical FFs:

They are defined as
\begin{align}\label{quarkdynFFdef}
 \Delta^\alpha_{F}(z,z_1) &=\frac{1}{N}\sum_X \iint
\frac{\drm \lambda d\mu}{(2\pi)^2}
\Exp{-\irm  \frac{\lambda}{z_1}}
\Exp{-i \mu \left( \frac{1}{z}-\frac{1}{z_1}\right)}
\bra{0} \psi (0) \ket{\Lambda^\uparrow X}
\bra{\Lambda^\uparrow X}\bar \psi (\lambda w)
g F^{\alpha w}(\mu w)\ket{0}  {\nonumber}\\
 & =M_h \epsilon^{\alpha S_\perp w P_h} \Slash{P}_h
\frac{\widehat D^*_{FT}(z,z_1)}{z}
-iM_h S_\perp^\alpha \gamma_5 \Slash{P}_h
\frac{\widehat G^*_{FT}(z,z_1)}{z} +\cdots,
\end{align}
$\widehat D_{FT}(z,z_1)$ and $\widehat G_{FT}(z,z_1)$ are complex functions
with the asterisk ${}^*$ denoting the complex conjugate.  
Their real parts are T-even, while their imaginary parts are
T-odd and contribute to $ep \to e\Lambda^\uparrow X$.

\subsubsection{Twist-3 gluon FFs}

Twist-3 gluon fragmentation functions can be also classified into three 
types\,\cite{Koike:2019zxc,Gamberg:2018fwy,Mulders:2000sh}

\vspace{3mm}

\noindent
(1) Intrinsic FFs: 

Intrinsic fragmentation functions are defined from
the lightcone correlation functions of the
field strength $F^{\mu \nu}_a$: 
\begin{align}\label{gluonintFFdef}
 \widehat\Gamma^{\alpha \beta}(z) &= \frac{1}{N^2-1} \sum_{X}
\int \frac{\drm \lambda}{2\pi} \Exp{-\irm \frac{\lambda}{z}}
\bra{0}F^{w \beta}_a(0)
\ket{\Lambda^\uparrow X}\bra{\Lambda^\uparrow X}
F^{w \alpha}_a(\lambda w) \ket{0}{\nonumber}\\
&= \irm M_h \epsilon^{\alpha \beta  w S_\perp }
\Delta \widehat G_{3T}(z)
+ M_h \epsilon^{P_h w S_\perp \{\alpha}w^{\beta\}}\Delta\widehat G_{3\bar T}(z)
+\cdots,
\end{align}
where 
$g^{\alpha\beta}_{\perp} = g^{\alpha\beta}- P_h^\alpha 
w^\beta - P_h^\beta w^\alpha$ and 
$\{ \alpha \beta\}$ indicates the symmetrization of the Lorentz indices.  
$\Delta \widehat G_{3T}(z)$ and $\Delta \widehat G_{3\bar T}(z)$ are
the intrinsic twist-3 fragmentation functions.  
$\Delta \widehat G_{3\bar T}(z)$ is T-odd and
contributes to the hyperon polarization.  

\vspace{3mm}

\noindent
(2) Kinematical FFs:

Kinematical fragmentation functions are defined from the correlation functions
containing the derivative of the field strengths: 
\begin{align}\label{gluonkinFFdef}
& \widehat \Gamma^{\alpha \beta \gamma}_\partial (z) =
\frac{1}{N^2 -1}\sum_{X} \int \frac{\drm \lambda}{2 \pi}
\Exp{- \irm \frac{\lambda}{z}} 
\bra{0}F^{w \beta}_a(0) \ket{\Lambda^\uparrow X}\bra{\Lambda^\uparrow X}
F^{w \alpha}_a(\lambda w)
\ket{0}\lpartial{}^\gamma {\nonumber}\\
 &=  -\irm \frac{M_h}{2}g_\perp^{\alpha \beta}
\epsilon^{P_h w S_\perp \gamma}\widehat G_T^{(1)}(z)
+ \frac{M_h}{2}\epsilon^{P_h w \alpha \beta}S_\perp^{\gamma}
\Delta \widehat G^{(1)}_T (z){\nonumber}\\
 &  -\irm \frac{M_h}{8}\left(
\epsilon^{P_h w S_\perp \{\alpha}g_\perp^{\beta \} \gamma}
+\epsilon^{P_h w \gamma \{\alpha}S^{\beta \}}_\perp
\right) \Delta \widehat H^{(1)}_T (z) +\cdots ,
\end{align}
where 
$\Delta \widehat G^{(1)}_T(z)$ is T-even, while 
$\widehat G^{(1)}_T(z)$ and $\Delta \widehat H^{(1)}_T(z)$ are T-odd
contributing to the hyperon polarization, and can be written as the 
$k^2_T/M^2_h$-moment of the TMD gluon fragmentation functions\cite{Mulders:2000sh}.  

\vspace{3mm}

\noindent
(3) Dynamical FFs:

Dynamical gluon FFs are defined from the correlation 
functions of three field-strengths.  Two types of FFs
exist, depending on the 
contraction of the color indices by two structure constants of color SU(N)
$-if^{abc}$, $d^{abc}$: 
\begin{align}\label{gluondynFAFFdef}
& \widehat\Gamma^{\alpha \beta \gamma}_{FA}
\left( \frac{1}{z_1},\frac{1}{z_2} \right) 
= \frac{- \irm f_{abc}}{N^2-1}\sum_{X}\iint
\frac{\drm \lambda d\mu}{(2 \pi)^2}
\Exp{-\irm  \frac{\lambda}{z_1}}
\Exp{- \irm \mu \left(\frac{1}{z_2}-\frac{1}{z_1}\right)}
\bra{0}F^{w \beta}_b(0)\ket{\Lambda^\uparrow X}
\bra{\Lambda^\uparrow X}F^{w \alpha}_a(\lambda w) gF^{w \gamma}_c(\mu w) \ket{0} {\nonumber}
\\
 &= -M_h \left(
\widehat N_1 \left( \frac{1}{z_1}, \frac{1}{z_2}\right)g^{\alpha \gamma}_\perp
\epsilon^{P_h w S_\perp \beta}
+\widehat N_2 \left( \frac{1}{z_1},\frac{1}{z_2} \right) g^{\beta \gamma}_\perp
\epsilon^{P_h w S_\perp \alpha}
 -\widehat N_2 \left(\frac{1}{z_2} -\frac{1}{z_1},\frac{1}{z_2} \right)
g^{\alpha \beta}_{\perp}\epsilon^{P_h w S_\perp \gamma}
\right) ,
\end{align}
and 
\begin{align}\label{gluondynFSFFdef}
 &\widehat\Gamma^{\alpha\beta\gamma}_{FS}
\left(\frac{1}{z_1}, \frac{1}{z_2}\right) 
 =\frac{d_{abc}}{N^2-1}\sum_{X}\iint\frac{\drm \lambda d\mu}{(2\pi)^2}
\Exp{-\irm  \frac{\lambda}{z_1}}
\Exp{-\irm \mu \left(\frac{1}{z_2}- \frac{1}{z_1}\right)}
\bra{0}F^{w \beta}_b(0) \ket{\Lambda^\uparrow X}
\bra{\Lambda^\uparrow X}F^{w\alpha}_a(\lambda w)
F^{w\gamma}_c(\mu w) \ket{0}{\nonumber}\\
 &=  -M_h \left(\widehat O_1 \left(\frac{1}{z_1},\frac{1}{z_2} \right)
g^{\alpha \gamma}_\perp \epsilon^{P_h w S_\perp \beta}
+ \widehat O_2 \left( \frac{1}{z_1},\frac{1}{z_2} \right)
g^{\beta \gamma}_\perp \epsilon^{P_h w S_\perp \alpha} 
 +\widehat O_2 \left(\frac{1}{z_2}- \frac{1}{z_1},\frac{1}{z_2} \right)
g^{\alpha \beta}_\perp \epsilon^{P_h w S_\perp \gamma}\right), 
\end{align}
where the support of each function is $1/z_2>1$ and $ 1/z_2 > 1/z_1 >0$.  
Four functions, 
$\widehat N_{1,2} \left(\frac{1}{z_1},\frac{1}{z_2}\right)$ and 
$\widehat O_{1,2} \left(\frac{1}{z_1},\frac{1}{z_2}\right)$, 
are complex functions and their imaginary parts are T-odd, 
contributing to SSAs,  and satisfies the following symmetry properties:  
\begin{align}\label{sym-property1}
 &\widehat N_1 \left(\frac{1}{z_1},\frac{1}{z_2} \right) =
 - \widehat N_1 \left( \frac{1}{z_2}- \frac{1}{z_1} ,\frac{1}{z_2}\right), &
& \widehat O_1 \left( \frac{1}{z_1}, \frac{1}{z_2}\right) 
= \widehat O_1 \left(\frac{1}{z_2} -\frac{1}{z_1} , \frac{1}{z_2}\right).&
\end{align}

For the complete description of the hyperon polarization,
another type of dynamical FFs constructed from the gluon's
field strength and the quark and anti-quark correlation functions are necessary: 
\begin{eqnarray}\label{gluondynFFdefanother}
 \widetilde\Delta^\alpha \left(\frac{1}{z_1},\frac{1}{z_2}\right)
&=&\frac{1}{N}\sum_{X}\iint \frac{\drm \lambda d\mu}{(2\pi)^2}
\Exp{-\irm  \frac{\lambda}{z_1}}
\Exp{-\irm\mu \left( \frac{1}{z_2}-\frac{1}{z_1}\right)}
\bra{0}F^{w\alpha}_a(\mu w) \ket{\Lambda^\uparrow X}\bra{\Lambda^\uparrow X}
g \bar \psi (\lambda w)t^a\psi(0)\ket{0}{\nonumber}\\
 &= &M_h \left(
\epsilon^{\alpha P_h w S_\perp} \Slash{P}_h 
\widetilde D_{FT} \left( \frac{1}{z_1},\frac{1}{z_2} \right)
+ \irm S^{\alpha }_\perp \gamma_5 \Slash{P}_h
\widetilde G_{FT} \left( \frac{1}{z_1}, \frac{1}{z_2} \right)
\right) .
\end{eqnarray}
The support of these functions is 
$1/z_1 >0$, $1/z_2 <0$ and $1/z_1 -1/z_2 >1$.  They are complex functions, 
imaginary parts of which are T-odd, contributing to SSAs.  
We remark that, 
although the above correlation function $\widetilde\Delta^\alpha$
contains quark and anti-quark fields like $\Delta_F^\alpha$ in (\ref{quarkdynFFdef}),
it is more appropriate to classify
$\widetilde D_{FT}$ and $\widetilde G_{FT}$ as the twist-3 dynamical
{\it gluon} FFs\cite{Ikarashi:2022yzg,Koike:2019zxc}.  

\subsubsection{Relations among the twist-3 FFs}

The twist-3 FFs introduced in the previous subsections obey the 
constraint relations among themselves, 
which are known as the QCD equation-of-motion (EOM) relations 
and the Lorentz invariance relations (LIRs). 
(See \cite{Kanazawa:2015ajw,Koike:2019zxc} for the derivation of 
the complete sets of
those relations for the twist-3 quark and gluon FFs.)
They play crucial roles to guarantee the gauge and frame-independence 
of various twist-3 cross 
sections including 
$pp\to \Lambda^\uparrow X$
\cite{Kanazawa:2015ajw,Koike:2017fxr,Koike:2021awj}.  
Here we quote those relations relevant to the present study from 
\cite{Kanazawa:2015ajw,Koike:2019zxc}, since
they are also indispensable to derive the consistency 
between the TMD factorization and 
the collinear twist-3 factorization for 
$ep\to e\Lambda^\uparrow X$.

For the T-odd twist-3 quark FFs, the EOM relation reads (eq. (22) of \cite{Kanazawa:2015ajw})
\begin{align}\label{EOMq}
 \int_{z}^{\infty} \frac{\drm z_1}{z_1^2}\frac{1}{1/z-1/z_1}
\left(
\Im \widehat D_{FT}(z,z_1) - \Im \widehat G_{FT}(z,z_1)
\right)
= \frac{D_T(z)}{z} + D_{1T}^{\perp (1)}(z), 
\end{align}
and the LIR reads (eq. (42) of \cite{Kanazawa:2015ajw})
\begin{align}\label{LIRq}
 -\frac{2}{z} \int_z^\infty \frac{\drm z_1}{z_1^2}
\frac{\Im \widehat D_{FT}(z,z_1)}{(1/z_1 - 1/z)^2}
= \frac{D_T(z)}{z} + \frac{\partial }{\partial (1/z)}\frac{D_{1T}^{\perp (1)}(z)}{z}. 
\end{align}

For the
T-odd twist-3 gluon FFs, EOM relation reads (eq. (50) of \cite{Koike:2019zxc})
\begin{align}\label{EOMg1}
 &\frac{1}{z}\Delta\widehat G_{3 \bar T}(z)
+ \Im \widetilde D_{FT}(z) - 
\frac{1}{2}\left(\widehat G_T^{(1)}(z)+\Delta\widehat H_T^{(1)}(z) \right)
{\nonumber}\\
 &= \int \drm \left(\frac{1}{z'}\right)\frac{1}{1/z -1/z'} \Im
\left(
2 \widehat N_1\left( \frac{1}{z'} ,\frac{1}{z}\right)
 + \widehat N_2\left( \frac{1}{z'},\frac{1}{z}\right)
-\widehat N_2 \left( \frac{1}{z}-\frac{1}{z'},\frac{1}{z}\right)
\right),
\end{align}
where $\widetilde D_{FT}(z)$ is defined as
\begin{align}
& \widetilde D_{FT}(z) \equiv
\frac{2}{C_F}\int_{0}^{1/z}\drm \left( \frac{1}{z'} \right)
\widetilde D_{FT}\left( \frac{1}{z'},\frac{1}{z'}-\frac{1}{z}\right),&
& C_F=\frac{N^2-1}{2N}.&
\end{align}
The LIRs combined with another EOM relation 
(as obtained from eqs. (53), (63) and (68) of \cite{Koike:2019zxc}) read
\begin{eqnarray}\label{EOMLIRg1}
 &&\frac{1}{z}\frac{\partial\widehat G_T^{(1)}(z)}{\partial(1/z)}
+2\left( \Im \widetilde D_{FT}(z)- \widehat G_T^{(1)}(z)\right){\nonumber}\\
 &=&4\int\drm \left(\frac{1}{z'}\right)\frac{1}{1/z-1/z'}
\Im\left(
\widehat N_1\left( \frac{1}{z'},\frac{1}{z}\right) 
- \widehat N_2 \left( \frac{1}{z}-\frac{1}{z'},\frac{1}{z}\right)
\right){\nonumber}\\
&&+2\int\drm \left( \frac{1}{z'}\right)\frac{1/z}{(1/z-1/z')^2}
\Im\left(
\widehat N_1\left( \frac{1}{z'},\frac{1}{z}\right) 
+\widehat N_2 \left( \frac{1}{z'},\frac{1}{z}\right)
-2\widehat N_2 \left( \frac{1}{z}-\frac{1}{z'},\frac{1}{z}\right)
\right),
\end{eqnarray}
and
\begin{eqnarray}\label{EOMLIRg2}
&& \frac{1}{z}\frac{\partial\Delta\widehat H_T^{(1)}(z)}{\partial(1/z)}
+4\left(\Im\widetilde D_{FT}(z)-\Delta\widehat H_{T}^{(1)}(z)\right){\nonumber}\\
&=&8\int\drm \left( \frac{1}{z'}\right)\frac{1}{1/z-1/z'}
\Im\left(
\widehat N_1 \left( \frac{1}{z'},\frac{1}{z}\right) 
+\widehat N_2 \left(\frac{1}{z'},\frac{1}{z}\right)
\right){\nonumber}\\
&&+4\int\drm \left( \frac{1}{z'}\right)\frac{1/z}{(1/z-1/z')^2}
\Im\left(
\widehat N_1 \left( \frac{1}{z'},\frac{1}{z}\right)
+\widehat N_2 \left( \frac{1}{z'},\frac{1}{z}\right)
\right).
\end{eqnarray}
These relations allow one to rewrite the derivatives of the
kinematical FFs in terms of other FFs.  

\subsection{TMD polarizing FF}

The hyperon polarization at small-$P_T$ can be described in the framework 
of the TMD factorization\cite{Ji:2004xq,Ji:2004wu}.  
Relevant TMD FF is the polarizing FF defined by \cite{Mulders:1995dh}:
\begin{align}\label{TMDFFdef}
 \Delta (z, \vec k_\perp)&=
\sum_X \int \frac{d\xi^+ d^2 \vec \xi_\perp}{2z (2\pi)^3} e^{ik \cdot \xi}
\bra{0} \psi (\xi) \ket{\Lambda^\uparrow X}
\bra{\Lambda^\uparrow X} \bar \psi (0) \ket{0}\Bigr|_{\xi^-=0}  {\nonumber}\\
 & =\frac{1}{2M_h P_h^-}\epsilon^{\alpha P_h k_\perp S_\perp} \gamma_\alpha
D_{1T}^\perp (z, z^2 \vec k_\perp^2)
+\cdots,
\end{align}
where $z= P_h^- /k^-$ is the longitudinal momentum fraction
(for the hyperon moving in the $-z$ direction).  
The hyperon's transverse momentum $\vec P_T$ with respect to the quark's momentum
is related to the above
$\vec k_\perp$ (which is defined in the frame with $P_{hT}=0$) as
$\vec P_T = - z \vec k_\perp$.  
$D_{1T}^\perp (z_f,z_f^2 \vec k_\perp^2)$ is the T-odd
polarizing FF.  
The kinematical twist-3 quark FF $D_{1T}^{\perp (1)}$ in (\ref{quarkkinFFdef}) is the 
$\vec k_\perp^2/M_h^2$-moment of $D_{1T}^\perp$: 
\begin{align}\label{moment}
 D_{1T}^{\perp (1)} (z)= z^2 \int d^2 \vec k_\perp
\left( \frac{\vec k_\perp^2}{2M_h^2}\right) D_{1T}^{\perp}(z,z^2 \vec k_\perp^2).
\end{align}
$D_{1T}^\perp$ convoluted with the TMD unpolarized
distribution in the nucleon contributes to the hyperon polarization
in the TMD factorization.


\section{Twist-3 FF contribution to 
$ep\to e\Lambda^\uparrow X$ at $\Lambda_{\rm QCD} \ll P_T\ll Q$}

\subsection{Structure of the twist-3 cross section}

Using the fragmentation functions defined in the previous section, 
the twist-3 quark and gluon fragmentation function contribution
to $ep\to e\Lambda^\uparrow X$ was derived, respectively, 
in \cite{Koike:2022ddx} and \cite{Ikarashi:2022yzg}.  
Here we summarize their basic structures.  

\vspace{3mm}

\noindent
(1) Twist-3 quark FF contribution:
\begin{eqnarray}
&&\hspace{-0.7cm}
\frac{d^6\Delta\sigma^{\rm tw3-quark}}{dx_{bj}dQ^2dz_fdq_T^2d\phi d\chi}\nn\\
&&=\frac{\alpha_{em}^{2}\alpha_s(-M_h)}{16\pi^2x_{bj}^2S_{ep}^2Q^2}
\sum_{k=1,2,3,4,8,9}
\mathscr{A}_{k}(\phi-\chi)\mathcal{S}_{k}\int_{x_{\rm min}}^1\frac{dx}{x}
\sum_{f_1=q,G}e_q^2\,f_1(x)
\int_{z_{\rm min}}^1\frac{dz}{z}\delta\left(\frac{q_T^2}{Q^2}-\left(1-\frac{1}{\xh}\right)
\left(1-\frac{1}{\zh}\right)\right)\nn\\
&&\times\Biggl[
\frac{D_T(z)}{z}\hat{\sigma}^{k}_T-\left\{\frac{d}{d(1/z)}
\frac{D^{\perp(1)}_{1T}(z)}{z}\right\}
\hat{\sigma}^{\it k}_{\perp D} -D^{\perp(1)}_{1T}(z)\hat\sigma^{\it k}_\perp\nn\\
&&+\int\frac{dz'}{{z'}^2}{\rm P}\left(\frac{1}{1/z-1/z'}\right)\Biggl\{{\Im}\widehat{D}_{FT}(z,z')
\left[\frac{z'}{z}\hat{\sigma}_{DF1}^{k}+
\frac{1}{z}\left(\frac{1}{z'}-
\frac{1}{\left(1-q_T^2/Q^2\right)z_f}\right)^{-1}\hspace{-3mm}\hat{\sigma}_{DF2}^{k}\right]\nn\\
&&+{\Im}\widehat{G}_{FT}(z,z')\left[
\frac{z'}{z}\hat{\sigma}_{GF1}^{k}+\frac{1}{z}
\left(\frac{1}{z'}-\frac{1}{\left(1-q_T^2/Q^2\right)z_f}\right)^{-1}\hspace{-3mm}
\hat{\sigma}_{GF2}^{\it k}\right]\Biggr\}
\Biggr], 
\label{qFraXsecfinal}
\end{eqnarray}
where $\displaystyle\sum_{f_1=q,G}$
indicates the sum over the twist-2 unpolarized quark distributions $q(x)$'s for 
quark flavor $q=u, d,\bar{u},\cdots$, 
and the twist-2 unpolarized gluon distribution $G(x)$, with $e_q$ representing
the fractional electric charge for the quark or anti-quark.   
Note that the flavor indices
for the twist-3 quark FFs are implicit in the above equation.  
In (\ref{qFraXsecfinal}), the lower limits 
of $x$ and $z$ integrations are, respectively, given by $x_{\rm min} =x_{bj} 
\left(1 + \frac{z_f}{1-z_f}\frac{q_T^2}{Q^2}\right)$ and
$z_{\rm min} = z_f \left( 1 + \frac{x_{bj}}{1-x_{bj}}\frac{q_T^2}{Q^2}\right)$ and 
${\Im}\widehat{D}_{FT}$ etc. indicates the imaginary part of $\widehat{D}_{FT}$. 
$\mathscr{A}_k(\phi-\chi)$'s are given in (\ref{Aks}), and 
${\cal S}_k=\sin \Phi_S$ ($k=1,\cdots 4$) and ${\cal S}_{8,9}=\cos\Phi_S$.  
The partonic hard cross sections
$\hat{\sigma}^{k}_T$, $\hat{\sigma}^{\it k}_{\perp D}$, $\hat\sigma^{\it k}_\perp$,
$\hat{\sigma}_{DF1}^{k}$, $\hat{\sigma}_{DF2}^{k}$, $\hat{\sigma}_{GF1}^{k}$
and $\hat{\sigma}_{GF2}^{k}$ which, of course, depend on whether $f_1=q$ or $G$
are the functions of $\hat{x}\equiv x_{bj}/x$, $\hat{z}\equiv z_f/z$, $Q$ and $q_T$ and 
can be found in \cite{Koike:2022ddx}.  

\vspace{3mm}

\noindent
(2) Twist-3 gluon FF contribution: 
\begin{eqnarray}
 &&\hspace{-0.5cm}\frac{\drm^6 
 \Delta\sigma^{\rm tw3-gluon}}{\drm x_{bj}\drm Q^2 \drm z_f \drm q_\Trm^2
\drm \phi \drm \chi}  {\nonumber}\\[7pt]
&&= \frac{\alpha_{em}^2 \alpha_s M_h}{16\pi^2 x_{bj}^2 S_{ep}^2Q^2}
\sum_{k}
 \mathscr{A}_k(\phi - \chi) \mathcal{S}_k
\int^1_{x_{\rm min}} {dx\over x}
\int^1_{z_{\rm min}} {dz\over z} 
z^2 \sum_q e_q^2\, q(x)
\delta \left(\frac{q_T^2}{Q^2} - \left(1-\frac{1}{\hat x} \right)\left(1-\frac{1}{\hat z} \right) \right) 
 {\nonumber}\\[7pt]
&&\times\left\{
{\widehat G^{(1)}_T (z)}
{ \hat \sigma^k_G } 
+
 {\Delta \widehat H^{(1)}_T (z)}
{\hat \sigma^k_H} \right. {\nonumber}\\[7pt]
&&+ \int\drm\left(\frac{1}{z'}\right)\left[
\frac{1}{1/z-1/z'}\Im \left(
{\widehat N_1 \left( \frac{1}{z'},\frac{1}{z} \right)}
{\hat\sigma^k_{N1}}
+
{\widehat N_2 \left( \frac{1}{z'},\frac{1}{z} \right)}
{\hat \sigma^k_{N2}}
+
{\widehat N_2 \left(\frac{1}{z}-\frac{1}{z'},\frac{1}{z} \right)}
{\hat \sigma^k_{N3}}
\right)
\right. {\nonumber}\\[7pt]
 &&+ \frac{1}{z}\left(\frac{1}{1/z-1/z'}\right)^2 \Im\left(
{\widehat N_1\left( \frac{1}{z'},\frac{1}{z} \right)}
{\hat \sigma^k_{DN1}}
+
 {\widehat N_2 \left( \frac{1}{z'},\frac{1}{z} \right)}
{\hat \sigma^k_{DN2}}
+ 
{\widehat N_2 \left(\frac{1}{z}-\frac{1}{z'},\frac{1}{z} \right)}
{\hat \sigma^k_{DN3}}
\right) {\nonumber}\\[7pt]
&&+ 
\frac{1}{1/z-1/z'}\Im \left(
{\widehat O_1 \left( \frac{1}{z'},\frac{1}{z} \right)}
{\hat\sigma^k_{O1}}
+
{\widehat O_2 \left( \frac{1}{z'},\frac{1}{z} \right)}
{\hat \sigma^k_{O2}}
+
{\widehat O_2 \left(\frac{1}{z}-\frac{1}{z'},\frac{1}{z} \right)}
{\hat \sigma^k_{O3}}
\right){\nonumber}\\[7pt]
&&\left. + \frac{1}{z}\left(\frac{1}{1/z-1/z'}\right)^2 \Im\left(
{\widehat O_1\left( \frac{1}{z'},\frac{1}{z} \right)}
{\hat \sigma^k_{DO1}}
+
 {\widehat O_2 \left( \frac{1}{z'},\frac{1}{z} \right)}
{\hat \sigma^k_{DO2}}
+ 
{\widehat O_2 \left(\frac{1}{z}-\frac{1}{z'},\frac{1}{z} \right)}
{\hat \sigma^k_{DO3}}
\right)\right] {\nonumber}\\[7pt]
&&
+ \int\drm\left(\frac{1}{z'}\right)\frac{2}{C_F} \left[ 
\Im 
{\widetilde D_{FT}\left(\frac{1}{z'},\frac{1}{z'}-\frac{1}{z} \right)} \right.
\left( 
{\hat \sigma^{k}_{DF1}}
+ \frac{1}{z}\frac{1}{1/z-1/z'}
{\hat \sigma^k_{DF2}}
+ \frac{z'}{z}
{\hat \sigma^k_{DF3}}  \right.  {\nonumber}\\[7pt]
&& \left. + \frac{1}{1-(1-q_T^2 / Q^2)z_f/z'}
{\hat \sigma^k_{DF4}}
+ \frac{1}{1-(1-q_T^2/Q^2)z_f(1/z-1/z')}
{\hat \sigma^k_{DF5}}
 \right) 
{\nonumber} \\[7pt]
&&+\Im 
{\widetilde G_{FT}\left(\frac{1}{z'},\frac{1}{z'}-\frac{1}{z} \right)} 
\left( 
\frac{1}{z}\frac{1}{1/z-1/z'}
{\hat \sigma^k_{GF2}}
+ \frac{z'}{z}
{\hat \sigma^k_{GF3}} \right.  {\nonumber}\\[7pt]
&&\left.   \left. \left. + \frac{1}{1-(1-q_T^2 / Q^2)z_f/z'}
{\hat \sigma^k_{GF4}}
+ \frac{1}{1-(1-q_T^2/Q^2)z_f(1/z-1/z')}
{\hat \sigma^k_{GF5}}
 \right)
 \right]
\right\}, 
\label{result}
\end{eqnarray}
where the flavor indices for $\widetilde D_{FT}$ and $\widetilde G_{FT}$ are again implicit
and each partonic hard cross section is given in \cite{Ikarashi:2022yzg}.  
The total twist-3 fragmentation function contribution to $ep\to e\Lambda^{\uparrow}X$
is the sum of (\ref{qFraXsecfinal}) and (\ref{result}), and is
decomposed into the five structure functions based on the dependence on the azimuthal 
angles as 
\begin{eqnarray}
 \frac{d^6 \Delta\sigma^{\rm tw3}}{d x_{bj} dQ^2 dz_f dq_T^2 d \phi d\chi}
 &=&\frac{d^6 \Delta\sigma^{\rm tw3-quark}}{d x_{bj} dQ^2 dz_f dq_T^2 d \phi d\chi}
 +\frac{d^6 \Delta\sigma^{\rm tw3-gluon}}{d x_{bj} dQ^2 dz_f dq_T^2 d \phi d\chi}\nonumber\\[7pt]
&=&\mathcal{F}_1 \sin \Phi_S
+ \mathcal{F}_2 \sin \Phi_S \cos (\phi - \chi)
+ \mathcal{F}_3 \sin \Phi_S \cos 2 (\phi -\chi) {\nonumber}\\[5pt]
& &+ \mathcal{F}_4 \cos \Phi_S \sin (\phi -\chi)
+ \mathcal{F}_5 \cos \Phi_S \sin 2 (\phi - \chi).
\label{tw3XsecAzimuth}
\end{eqnarray}

\subsection{Twist-3 cross section at $\Lambda_{\rm QCD}\ll P_T \ll Q$}

In order to investigate the consistency between the twist-3 cross section (\ref{tw3XsecAzimuth}) and
the corresponding TMD factorization formula, one needs to examine
the behavior of  (\ref{tw3XsecAzimuth}) in the region of the small
transverse momentum $P_T=z_fq_T \ll Q$.  
To this end, we note 
\begin{align}
\left.\delta \left(
\frac{q_T^2}{Q^2} - \left( 1 - \frac{1}{\hat x}\right)
\left( 1 - \frac{1}{\hat z}\right)
       \right)\right|_{q_T\ll Q}
=
\hat x \hat z \left(
\frac{\delta(\hat x - 1)}{(1 - \hat z )_+}
+ \frac{\delta(\hat z - 1)}{(1 - \hat x )_+}
+ \delta( \hat x - 1) \delta( \hat z - 1) \ln \frac{Q^2}{q_T^2}
\right), 
\end{align}
where $+$-function is defined as
\begin{align}
 \int_z^1 dx\ T(x) \frac{1}{(1-x)_+} \equiv
\int_z^1 dx \frac{T(x)-T(1)}{1-x} + T(1) \ln (1-z)
\end{align}
for an arbitrary function $T(x)$.  
Using the actual form of the partonic hard cross sections, one finds that 
only $\mathcal{F}_1$ in (\ref{tw3XsecAzimuth})  receives the leading 
contribution with respect to ${q_T/Q}$: 
\begin{align}\label{F1}
\left.\frac{d^6 \Delta\sigma^{\rm tw3}}{d x_{bj} dQ^2 dz_f dq_T^2 
d \phi d\chi}\right|_{q_T\ll Q} =\left.\mathcal{F}_1\right|_{q_T\ll Q}\sin \Phi_S, 
\end{align}
with
\begin{align}\label{F1lim}
 \left.\mathcal{F}_1\right|_{q_T\ll Q} &=
\frac{-4 \alpha_s M_h \sigma_0}{(2\pi)^2 q_T^3}\sum_q e_q^2
\left[
q (x_{bj}) \int \frac{dz}{z} \left( A - \frac{1}{4}B\right)
+ C_F D_{1T}^{\perp (1)} (z_f)
\int \frac{dx}{x}q(x)
\frac{1+ \hat x^2}{(1- \hat x)_+}  \right. {\nonumber}\\
& \left. 
+ \frac{1}{2}D_{1T}^{\perp (1)}(z_f) \int \frac{dx}{x}G(x)(1-2\hat x + 2\hat x^2)
+ 2C_F\, q (x_{bj}) D_{1T}^{\perp (1)}(z_f) \ln \frac{Q^2}{q_T^2}
\right], 
\end{align}
where $\sigma_0=\alpha_{\rm em}^2(1-y+y^2/2)/Q^4$ with 
$y\equiv p\cdot q/p\cdot\ell=Q^2/(x_{bj}S_{ep})$. 
$A$ and $B$ in (\ref{F1lim}) are, respectively, the twist-3
quark and gluon FF contributions and are given by
\footnote{Note that $A$ and $B$ also carry the flavor indices from twist-3 FFs, 
which are suppressed here.}
\begin{align}\label{A}
A&=
\frac{D_T (z)}{z}\left(
-C_F (1+2\hat z) - \frac{1}{2 N}\frac{1+\hat z^2}{\hat z}
\right) {\nonumber} \\
& 
+  \left(\frac{\partial}{\partial (1/z)}\frac{D_{1T}^{\perp (1)}(z)}{z}
\right)
\left(
- \frac{1}{2 N} \frac{1+\hat z^2}{\hat z}
\right)
+ D_{1T}^{\perp (1)}(z) \left(
C_F \frac{\hat z( 1+\hat z)}{(1-\hat z)_+}
\right) {\nonumber}\\
  & 
+ \int d \left(\frac{1}{z'}\right) \frac{1/z}{1/z-1/z'} 
\Im \widehat D_{FT}(z,z')\left\{
\frac{1}{1/z'}\frac{1}{2N}\frac{2-\hat z}{\hat z}
+ \frac{1}{1/z'-1/z_f}\left(C_F +\frac{1}{2N}\right)(1+\hat z)
\right\} {\nonumber}\\
 & 
+ \int d \left(\frac{1}{z'}\right) \frac{1/z}{1/z-1/z'} 
\Im \widehat G_{FT}(z,z')\left\{
\frac{1}{1/z'}\frac{1}{2N}
- \frac{1}{1/z'-1/z_f}\left(C_F +\frac{1}{2N}\right)(1-\hat z)
\right\},
\end{align}
and
\begin{align}\label{B}
B &= 2C_F z^2 \left[
 \frac{(1 - \hat z) (-2 + \hat z^2)}{\hat z^2}
\widehat G_T^{(1)}(z)
- 2 \frac{(1 - \hat z)}{\hat z}
\Delta \widehat H_T^{(1)}(z)
+ \int d\left(\frac{1}{z'}\right) \frac{1}{1/z - 1/z'}
\frac{(1 - \hat z)}{\hat z^2}
\right.  {\nonumber} \\
& \left. \times
\Im \left\{
 4(2 -3\hat z +\hat z^2)
\widehat N_1\left(\frac{1}{z'},\frac{1}{z}\right)
+2(2 - 3\hat z)
\widehat N_2\left(\frac{1}{z'},\frac{1}{z}\right)
-2(2 - 3\hat z +2 \hat z^2)
\widehat N_2 \left(\frac{1}{z}-\frac{1}{z'},\frac{1}{z}\right)
\right\}
\right. {\nonumber} \\
& \left. +
\int d \left(\frac{1}{z'}\right)
\frac{1}{z} \left(\frac{1}{1/z - 1/z'}\right)^2
\frac{(1 - \hat z)}{\hat z^2}
\right.  {\nonumber}\\
&\left. \times
\Im \left\{
(4-4\hat z + \hat z^2)
\widehat N_1\left(\frac{1}{z'},\frac{1}{z}\right)
+(4-4\hat z + \hat z^2)
\widehat N_2\left(\frac{1}{z'},\frac{1}{z}\right)
-2(2- 2\hat z + \hat z^2) 
\widehat N_2\left(\frac{1}{z}-\frac{1}{z'},\frac{1}{z} \right)
\right\}
\right.  {\nonumber}\\
&\left. +\int d\left(\frac{1}{z'}\right)
\frac{1}{1/z-1/z'} \frac{(1 - \hat z)}{\hat z^2}
\right.  {\nonumber}\\
&\left. \times
\Im \left\{
4 (2 - \hat z) \widehat O_1\left(\frac{1}{z'},\frac{1}{z}\right)
+2(4-3\hat z + \hat z^2)
\widehat O_2\left(\frac{1}{z'},\frac{1}{z}\right)
+2(4-3\hat z + \hat z^2)
\widehat O_2\left(\frac{1}{z}-\frac{1}{z'},\frac{1}{z}\right)
\right\}
\right.  {\nonumber}\\
&\left. + \int d\left(\frac{1}{z'}\right)
\frac{1}{z}\left(\frac{1}{1/z-1/z'}\right)^2
\frac{(1 - \hat z)}{\hat z^2}
\right.  {\nonumber}\\
&\left. \times \Im
\left\{
(4-4\hat z + \hat z^2)\widehat O_1\left(\frac{1}{z'},\frac{1}{z}\right)
+(4-4\hat z + \hat z^2)\widehat O_2\left(\frac{1}{z'},\frac{1}{z}\right)
+2(2-2\hat z+ \hat z^2)
\widehat O_2\left(\frac{1}{z}-\frac{1}{z'},\frac{1}{z}\right)
\right\}
\right.  {\nonumber}\\
&\left. +
\frac{1}{C_F} \int d\left(\frac{1}{z'}\right) \Im 
\widetilde D_{FT}\left(\frac{1}{z'},\frac{1}{z'}-\frac{1}{z}\right)
\right.  {\nonumber}\\
&\left. \times
\left\{
-4\frac{(1 - \hat z)}{\hat z^2}(2-3\hat z +\hat z^2)
+\frac{1}{N}\frac{1}{z}\frac{1}{1/z-1/z'}
\frac{-1}{\hat z}(2-\hat z)
+\frac{1}{N}\frac{1}{1-z_f / z'}
(-\hat z)(-2 + \hat z)
\right\}
\right.  {\nonumber}\\
&\left. +
\frac{1}{C_F}\int d\left(\frac{1}{z'}\right)
\Im \widetilde G_{FT}\left(\frac{1}{z'},\frac{1}{z'}-\frac{1}{z}\right)
\left\{
\frac{1}{N}\frac{1/z}{1/z-1/z'}
+\frac{1}{N}\frac{1}{1- z_f / z'}
(-\hat z^2)
\right\}
\right].
\end{align}
We note that, at $P_T/Q\to 0$, $\vec P_h$ and $\vec q$ become parallel 
and the azimuthal angle $\Phi_S$
can be regarded as the angle around $\vec q$ as is seen from Fig. \ref{hadronframe}.

\section{TMD factorization for $ep\to e\Lambda^\uparrow X$ 
at $P_{T}\gg\Lambda_{\rm QCD}$}

\subsection{Framework}

When the transverse momentum of the hyperon $P_T$
is much smaller than the hard 
scale, i.e., $\Lambda_{\mathrm{QCD}}\lesssim P_{T}=z_f q_T \ll Q$, 
the TMD factorization can be applied to $ep\to e\Lambda^\uparrow X$. 
In this framework, the $\sin\Phi_S$ asymmetry can be described in terms of the 
polarizing FF $D_{1T}^\perp$\cite{Mulders:1995dh,Ji:2004xq,Ji:2004wu} as
\begin{align}\label{TMDstart}
 \frac{d^6 \sigma}{d x_{bj} dQ^2 dz_f dq_T^2 d \phi d\chi}
&=-z_f^2 \sigma_0 \sin \Phi_S
\int d^2 \vec k_\perp d^2 \vec p_\perp d^2 \vec \lambda_\perp
\delta^2 (\vec k_\perp - \vec p_\perp + \vec \lambda_\perp 
- \vec P_{T}/z_f)
{\nonumber} \\
&\times
\frac{\vec p_\perp^2}{q_T M_h}\sum_q e_q^2\,
q(x_{bj}, k_\perp^2 )
D_{1T}^{\perp}(z_f, z_f^2 p_\perp^2) S^{-1}(\lambda_\perp^2) H(Q^2),
\end{align}
where $q(x_{bj},k_\perp^2)$ is the TMD unpolarized quark DF, 
$H(Q^2)$ is the hard factor which is calculable in perturbative QCD and takes the form of
$H(Q^2)=1+\mathcal O(\alpha_s)$, 
and the soft factor $S(\lambda_\perp^2)$ is the vacuum expectation value of the Wilson line
representing the soft gluon radiation with 
$S^{-1}$ canceling the effect of the soft gluon radiation in the TMD functions.  
In order to see the consistency with the collinear twist-3 factorization,
we see the behavior of (\ref{TMDstart}) in large-$q_T$ region, 
$q_T\gg\Lambda_{\mathrm{QCD}}$.  
The dominant contribution of
(\ref{TMDstart}) in this region is obtained
by setting one of $\vec k_\perp$, $-\vec p_\perp$ and $\vec \lambda_\perp$ in the 
$\delta$-function equal to 
$\vec P_{T}/z_f$ with other two transverse momenta set to zero.  
Then (\ref{TMDstart}) becomes
\begin{align}\label{TMDcrosssection}
 \frac{d^6 \sigma}{d x_{bj} dQ^2 dz_f dq_T^2 d \phi d\chi} 
&=-2 \sigma_0 \sin \Phi_S \frac{M_h}{q_T}\sum_q e_q^2
\left(
q(x_{bj}, q_T^2 ) D_{1T}^{\perp (1)}(z_f)
\right. {\nonumber}\\
&\left.
+ \frac{P_{T}^2}{2M_h^2}q(x_{bj})D_{1T}^{\perp}(z_f, P_{T}^2)
+q(x_{bj})
D_{1T}^{\perp (1)}(z_f) S^{-1}(q_T^2) 
\right), 
\end{align}
where 
we have used the following relation, 
\begin{align}
 \int d^2 \vec k_\perp q(x_{bj},\vec k_\perp^2)= q(x_{bj}), 
\end{align}
and 
$D_{1T}^{\perp (1)}(z_f)$ is the twist-3 quark FF obtained as the $p_\perp^2/M_h^2$-moment
of the TMD polarizing FF:  
\begin{align}
 D_{1T}^{\perp (1)}(z_f)
= z_f^2 \int d^2 p_\perp \frac{\vec p_\perp^2}{2M_h^2}
D_{1T}^{\perp}(z_f,z_f^2 p_\perp^2).
\end{align}
Since we calculate $q(x_{bj},q_T^2)$, $D_{1T}^{\perp}(z_f,P_{T}^2)$, $S^{-1}(q_T^2)$ 
in the LO with respect to the QCD coupling constant, 
the hard factor is set equal to one.  
In order to investigate the behavior of (\ref{TMDcrosssection}) at
$\Lambda_{\mathrm{QCD}}\ll q_T \ll Q$, 
we utilize the fact that the $q_T$-dependence of each 
factor in (\ref{TMDcrosssection})
can be calculated by perturbative QCD\cite{Ji:2006ub,Ji:2006br,Ji:2006vf,Koike:2007dg}. 
A perturbative calculation of the TMD DFs and FFs 
at large 
intrinsic transverse momentum involves the rapidity divergence, if one uses
the lightcone vectors $n^\mu$ and $w^\mu$ ($n^2=w^2=0$). 
To regularize this divergence, we introduce non-lightlike vectors
$v^\mu$ and $\tilde v^\mu$:  
\begin{align}
&v = (v^+, v^-, \vec 0_\perp), & &\tilde v = (\tilde v^+, \tilde v^-, \vec 0_\perp), &
\end{align}
and define
\begin{align}
 \rho^2 =  \frac{(2 v \cdot \tilde v)^2}{v^2 \tilde v^2}. 
\end{align}
As $v^+ \to 0$, $\tilde v^- \to 0$, $v$ and $\tilde v$ become lightlike ($v^2 = \tilde v^2 =0$) and 
$\rho^2 \to \infty$.  
The soft factor has been obtained as\cite{Ji:2006br,Ji:2006vf}
\begin{align}\label{softinverse}
 S^{-1}(q_T^2) = - \frac{\alpha_s C_F}{2\pi^2 q_T^2}(\ln \rho^2 - 2). 
\end{align}
The transverse momentum dependence of the TMD functions can be calculated by perturbative QCD
at
$q_T \gg \Lambda_{\mathrm{QCD}}$, 
and they can be expressed in terms of the collinear PDFs and FFs.
The result for the unpolarized DF is well known\cite{Ji:2004wu,Ji:2006vf}: 
\begin{align}\label{unpolTMD}
 q(x_{bj}, q_T^2) &= 
\frac{\alpha_s C_F}{2\pi^2 q_T^2}
\int \frac{dx}{x} q(x) \left[
\frac{1 + \hat x^2}{(1-\hat x)_+}
+\delta (1-\hat x) \left(
\ln \frac{x_{bj}^2 \zeta^2}{ q_T^2} - 1
\right)
\right]
{\nonumber}\\
 &\
 + \frac{\alpha_s}{2\pi^2  q_T^2} \frac{1}{2}\int \frac{dx}{x} G(x)
\left(
1 - 2 \hat x + 2 \hat x^2
\right), 
\end{align}
where $\zeta^2 \equiv (2 v \cdot p)^2/ v^2$ ($\zeta \to \infty$ as $v^+ \to 0$) is the regulator for the rapidity divergence.  

\subsection{Perturbative calculation of the polarizing FF $D_{1T}^\perp (z,z^2k_\perp^2)$
at $k_\perp\gg \Lambda_{\rm QCD}$}

\subsubsection{Factorization of the TMD FF into the Twist-3 FFs and the hard part}

\begin{figure}[htb]
\centering
 \includegraphics[scale=1.0]{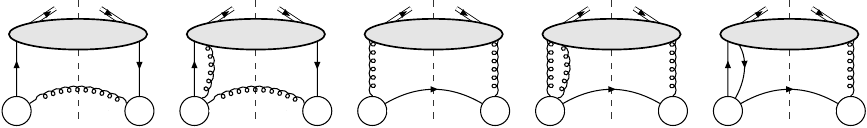}
\caption{Feynman diagrams contributing to $D_{1T}^\perp$
at $q_T \gg \Lambda_{\mathrm{QCD}}$.  
Mirror diagrams also contribute. 
From left to right, 
each diagram corresponds to $\Delta (\ell) S(\ell)$, $\Delta_{La}^\alpha (\ell,\ell')S_\alpha^{La}(\ell,\ell')$, 
$\widehat \Gamma_{ab}^{\mu\nu}(\ell) S_{\mu\nu}^{ab}(\ell)$, $\widehat \Gamma_{Labc}^{\mu\nu\lambda}(\ell,\ell')
S^{Labc}_{\mu\nu\lambda}(\ell,\ell')$, $\widetilde \Delta_{La}^\alpha (\ell,\ell') \widetilde S^{La}_{\alpha}(\ell,\ell')$.  
Upper portion in each diagram represents the fragmentation matrix elements (for the twist-3 FFs),
while the lower white circles correspond to the partonic hard part. 
In the second, fourth and fifth diagram, 
momenta of the parton lines in the left of the cut are $\ell$ and $\ell'-\ell$ flowing upwards.  }
\label{TMDgeneric}
\end{figure}

Next we consider the perturbative calculation of the polarizing FF $D_{1T}^\perp$
at 
$q_T \gg \Lambda_{\mathrm{QCD}}$.  Following the method for the
Sivers function and the Collins function\cite{Ji:2006br,Ji:2006vf,Yuan:2009dw}, 
one can factorize the correlation function for the TMD FF (\ref{TMDFFdef})
into the collinear twist-3 FFs and the hard part.  To this end, we decompose
(\ref{TMDFFdef}) into the quark and gluon pieces: 
\begin{align}\label{setup}
  \Tr{} \Delta(z_f,\vec k_\perp) \frac{\Slash{w}}{2} =
\left[ \Tr{} \Delta(z_f,\vec k_\perp) \frac{\Slash{w}}{2} \right]^{\mathrm{q}}
+
\left[ \Tr{} \Delta(z_f,\vec k_\perp) \frac{\Slash{w}}{2} \right]^{\mathrm{g}},
\end{align}
where the first and the second terms, respectively, correspond to
the quark- and gluon- fragmentation channels, and they are given as
\begin{align}\label{setupqchannel}
 \left[ \Tr{} \Delta(z_f,\vec k_\perp) \frac{\Slash{w}}{2} \right]^{\mathrm{q}}
\equiv \Tr{}\int \frac{d^4 \ell}{(2\pi)^4} \Delta^{}(\ell)S(\ell)
+ \Tr{}\iint \frac{d^4 \ell d^4 \ell'}{(2\pi)^8}
\Delta^{\alpha}_{La}(\ell,\ell')S^{La}_\alpha(\ell,\ell')
+ [L \rightarrow R].  
\end{align}
\begin{align}\label{setupgchannel}
\left[ \Tr{} \Delta(z_f,\vec k_\perp) \frac{\Slash{w}}{2} \right]^{\mathrm{g}}
&\equiv
 \Tr{} \int \frac{d^4 \ell}{(2\pi)^4} \widehat \Gamma^{\mu\nu}_{ab}(\ell)
 S_{\mu\nu}^{ab}(\ell)
+\frac{1}{2}\Tr{} \iint \frac{d^4 \ell d^4\ell'}{(2\pi)^8}
\widehat \Gamma^{\mu\nu\lambda}_{Labc}(\ell,\ell')S_{\mu\nu\lambda}^{Labc}(\ell,\ell')
{\nonumber}\\
&
+ \Tr{}\iint \frac{d^4 \ell d^4 \ell'}{(2\pi)^8}
\widetilde \Delta_{La}^{\alpha}(\ell,\ell')\widetilde S_{\alpha}^{La}(\ell,\ell')
+ [L \rightarrow R],
\end{align}
where $\Tr{}$ indicates sum over the spinor or Lorentz indices depending on channels.  
$D_{1T}^\perp$ is identified as the coefficient 
of $\epsilon^{\alpha P_h k_\perp S_\perp}\gamma_\alpha$ in 
$\Delta(z_f, \vec k_\perp)$:  
\begin{align}
  \Tr{} \Delta(z_f,\vec k_\perp) \frac{\Slash{w}}{2}
= \frac{\epsilon^{wP_h k_\perp S_\perp}}{M_h P_h^-}
D_{1T}^{\perp}(z_f, z_f^2 k_\perp^2) +\cdots, 
\end{align}
where we have dropped the terms corresponding to the unpolarized FF.  
Correspondingly, we pick up terms proportional to $\epsilon^{wP_h k_\perp S_\perp}$
in the RHS of (\ref{setup}).  
$\Delta(\ell)$ and $\Delta_{La}^{\alpha}(\ell,\ell')$ in (\ref{setupqchannel}) are
the fragmentation matrix elements corresponding, respectively, to 
$\mathcal{FT}\sum_X\bra{0}\psi \ket{\Lambda^\uparrow X}\bra{\Lambda^\uparrow X}
\bar \psi \ket{0}$ and 
$ 
\mathcal{FT}\sum_X\bra{0}\psi \ket{\Lambda^\uparrow X}\bra{\Lambda^\uparrow X}
\bar \psi g A^\alpha_a \ket{0}
$ as defined in Eqs. (18) and (19) of  \cite{Kanazawa:2013uia}.  
$\widehat \Gamma_{ab}^{\mu\nu}(\ell)$, $\widehat \Gamma_{Labc}^{\mu\nu\lambda}(\ell,\ell')$
and $\widetilde \Delta_{La}^{\alpha}(\ell,\ell')$ in (\ref{setupgchannel}) are
the gluon fragmenting matrix elements
corresponding, respectively, to
$\mathcal{FT}\sum_X\bra{0}A^\nu_b\ket{\Lambda^\uparrow X}\bra{\Lambda^\uparrow X}  A^\mu_a    \ket{0}$,\\
$\mathcal{FT}\sum_X\bra{0}A^\nu_b\ket{\Lambda^\uparrow X}\bra{\Lambda^\uparrow X}  A^\mu_a  gA^\lambda_c  \ket{0}$, 
and 
$\mathcal{FT}\sum_X\bra{0}gA^\alpha_a\ket{\Lambda^\uparrow X}\bra{\Lambda^\uparrow X}
\psi  \bar \psi\ket{0}$ as defined in Eqs. (32), (33) and (35) of \cite{Ikarashi:2022yzg}.  
$S(\ell)$, $S_\alpha^{La}(\ell,\ell')$, $S_{\mu\nu}^{ab}(\ell)$, $S_{\mu\nu\lambda}^{Labc}(\ell,\ell')$ and 
$\widetilde S_\alpha^{La}(\ell,\ell')$
are the partonic hard part before collinear expansion in each channel.  
The symbol 
$L$, ($R$) represents that two fragmenting parton lines reside in the left (right) of the cut, and 
$[L \to R]$ indicates the contribution from the mirror diagram.
Figure \ref{TMDgeneric}
shows the generic diagrams for the twist-3 FF contribution to
$D_{1T}^\perp$. 
These diagrams show the TMD FF in which a quark with momentum $(z_f, \vec k_\perp)$
splits into partons with momenta represented by $\ell$ (and/or $\ell'$) and eventually fragments into the polarized hyperon
through the twist-3 FF.   The mixing with purely gluonic channels also occurs.  
Each term in 
(\ref{setupqchannel}) and (\ref{setupgchannel})
is not gauge invariant, but the reorganization of the collinear expansion combined with the Ward-Takahashi identities
lead to a gauge-invariant twist-3 FF contribution to the process.  

Although the hard parts $S(\ell)$ and $S^{La}_\alpha (\ell, \ell')$ $\cdots$
contains the Wilson lines,  they
satisfy the following Ward-Takahashi identities
which are formally the same as in the case of the calculation of the twist-3
cross section\cite{Kanazawa:2013uia,Ikarashi:2022yzg}.
They read
\begin{align}
\label{WTidentity1}
&(\ell' - \ell)^\alpha S_\alpha^{La}(\ell, \ell')=T^a S(\ell') +G^a(\ell,\ell'),&
\end{align}
for the quark fragmentation channel, and 
\begin{align}
&\ell^\mu S_{\mu\nu}^{ab}(\ell)= \ell^\nu S_{\mu\nu}^{ab}(\ell) =0, & \\
&\ell^\mu S^{Labc}_{\mu\nu\lambda}(\ell,\ell')
= \frac{if^{abc}}{N^2-1}S_{\lambda\nu}(\ell'), & \\
&(\ell'-\ell)^\lambda S^{Labc}_{\mu\nu\lambda}(\ell,\ell')
=\frac{-if^{abc}}{N^2-1}S_{\mu\nu}(\ell'),& \\
& \ell'^\nu S^{Labc}_{\mu\nu\lambda}(\ell,\ell')=0, & \\
&(\ell - \ell')^\alpha \widetilde S^{La}_\alpha (\ell,\ell') = 0,& 
\label{WTidentity2}
\end{align}
for the gluon fragmentation channel with 
$S_{\lambda \nu} (\ell') \equiv  \delta^{ab}S_{\lambda\nu}^{ab}(\ell')$.  
Although the ghost-like terms $G^a(\ell,\ell')$ appear in the quark fragmentation channel, 
they eventually drop in the calculation\cite{Kanazawa:2013uia}.  
In the gluon fragmentation channel, such ghost-like terms do not appear 
as in the calculation of the twist-3 cross section\cite{Ikarashi:2022yzg}.  
Accordingly, 
(\ref{setupqchannel}) and (\ref{setupgchannel}) can be analyzed by using the same method 
used in the derivation of the twist-3 FF contribution to $ep\to e\Lambda^\uparrow X$ in
\cite{Koike:2022ddx,Ikarashi:2022yzg}.  

This way, we obtain from (\ref{setupqchannel}) the twist-3 quark FF contribution
to the polarizing FF in a gauge-invariant form as
\begin{align}\label{setupq}
\left[ \Tr{} \Delta(z_f,\vec k_\perp) \frac{\Slash{w}}{2} \right]^{\mathrm{q-tw3}}
&=
\int \frac{dz}{z^2}\left[
\Tr{} \Delta(z) S (P_h/z) -i \Tr{} \Omega^\alpha_{\; \beta}
\Delta^\beta_\partial (z) 
\left.\frac{\partial S(k)}{\partial k^\alpha}\right|_{\cl} 
\right.{\nonumber} \\
&\left. + 2 \Re\, i \int \frac{dz'}{z'^2} \frac{1}{1/z-1/z'}
\Tr{} \Omega^\alpha_{\; \beta}\Delta^\beta_F (z',z)
S_{L\alpha}(1/z',1/z)
\right],
\end{align}
where $\Omega^\alpha_{\ \beta}\equiv g^\alpha_{\ \beta}-P_h^\alpha w_\beta$, and
$\Delta$, $\Delta^\beta_\partial$ and $\Delta_F^\beta$ are, respectively, 
the correlation functions for the 
intrinsic, kinematical and dynamical twist-3 quark FFs given in 
(\ref{quarkintFFdef}), \ref{quarkkinFFdef}) and \ref{quarkdynFFdef}).  
Likewise from 
(\ref{setupgchannel}), one obtains the gauge-invariant twist-3 gluon FF contribution 
to the polarizing FF as 
\begin{align}\label{setupg}
\left[ \Tr{} \Delta(z_f,\vec k_\perp) \frac{\Slash{w}}{2} \right]^{\mathrm{g-tw3}}
 &= \Omega^\alpha_\mu \Omega^\beta_\nu
\int d\left(\frac{1}{z}\right) z^2 \widehat \Gamma^{\mu \nu}(z)S_{\alpha\beta}(1/z)
- i \Omega^\alpha_\mu \Omega^\beta_\nu \Omega^\gamma_\lambda
\int d\left(\frac{1}{z} \right) z^2 \widehat \Gamma^{\mu\nu\lambda}_\partial(z)
\left. \frac{\partial S_{\alpha\beta}(k)}{\partial k^\gamma}\right|_\cl{\nonumber} \\[5pt]
  & + \Re \left[
i \Omega^\alpha_\mu \Omega^\beta_\nu \Omega^\gamma_\lambda
\iint d \left(\frac{1}{z} \right)d \left(\frac{1}{z'}\right)\frac{zz'}{1/z-1/z'}
\right. {\nonumber}\\[5pt]
&\left.\times
\left\{
- \frac{if^{abc}}{N}
\widehat \Gamma^{\mu\nu\lambda}_{FA}\left( \frac{1}{z'},\frac{1}{z}\right)
+ \frac{Nd^{abc}}{N^2-4} 
\widehat\Gamma^{\mu\nu\lambda}_{FS}\left( \frac{1}{z'},\frac{1}{z}\right)
\right\}S^{Labc}_{\alpha\beta\gamma}\left( \frac{1}{z'},\frac{1}{z}\right)
\right] {\nonumber}\\[7pt]
  & + \Re \left[
i \Omega^\alpha_\mu \iint d\left(\frac{1}{z}\right)d\left(\frac{1}{z'}\right)z
\Tr{} \widetilde \Delta^\mu\left(\frac{1}{z'},\frac{1}{z'}-\frac{1}{z}\right) 
\widetilde S^L_\alpha \left( \frac{1}{z'},\frac{1}{z'}-\frac{1}{z}\right)
\right],
\end{align}
where $\widehat \Gamma^{\mu\nu}$, $\widehat \Gamma_\partial^{\mu\nu\lambda}$, 
$\widehat \Gamma_{FA}^{\mu\nu\lambda}$, 
$\widehat \Gamma_{FS}^{\mu\nu\lambda}$ and $\widetilde \Delta^\mu$
are given in (\ref{gluonintFFdef}), (\ref{gluonkinFFdef}), (\ref{gluondynFAFFdef}), 
(\ref{gluondynFSFFdef}) and \ref{gluondynFFdefanother}).  

\subsubsection{Quark fragmentation channel}

\begin{figure}[htb]
\begin{center}
\includegraphics[scale=1]{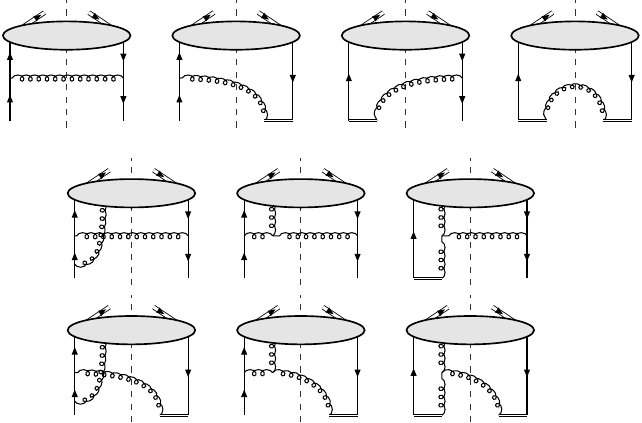}
\end{center}
 \caption{Twist-3 quark FF contribution to the TMD polarizing FF at large $k_\perp$.  
Diagrams in which a gluon line directly connects the twist-3 FF and the Wilson line
become zero, so that they are not shown. }
\label{diagquarkfrag}
\end{figure}

We first calculate the twist-3 quark FF contribution
to the TMD polarizing FF (\ref{setupq}).  
Fig. \ref{diagquarkfrag} shows the relevant diagrams.  
Some diagrams contain Wilson lines, which are absent from 
the hard parts for
the twist-3 quark FF contribution to the cross section. 

The contribution from $\Delta(z)$ (the first term in $[\ \ ]$ of (\ref{setupq}))
becomes
\begin{align}\label{quark-int}
\left[ \Tr{} \Delta(z_f,\vec k_\perp) \frac{\Slash{w}}{2} 
\right]^{\mathrm{q-tw3}}_{\Delta}=
\frac{4\alpha_s M_h C_F \epsilon^{k_\perp P_h S_\perp w}}
{ (2\pi)^2P_h^- \vec k_\perp^4 z_f^2}
\int \frac{dz}{z}\frac{D_T(z)}{z}\frac{1}{\hat z}.  
\end{align}

The contribution from $\Delta_\partial (z)$ (the second term in $[\ \ ]$ of (\ref{setupq}))
is obtained as
\begin{align}\label{quark-kin}
\left[ \Tr{} \Delta(z_f,\vec k_\perp) \frac{\Slash{w}}{2} 
\right]^{\mathrm{q-tw3}}_{\Delta_\partial}
& =
\frac{4\alpha_s M_h C_F \epsilon^{k_\perp P_h S_\perp w}}
{ (2\pi)^2P_h^- \vec k_\perp^4 z_f^2}
\int \frac{dz}{z}\left[
D_{1T}^{\perp (1)}(z) \left\{
\frac{1+\hat z}{(1-\hat z)_+} + \delta(1-\hat z) 
\left(
\ln \frac{\hat \zeta^2}{\vec k_\perp^2 z_f^2}
 - 1 \right)
\right\} \right.  {\nonumber}\\
&\left. + \left( \frac{\partial}{\partial (1/z)}\frac{D_{1T}^{\perp (1)}(z)}{z}
\right) \frac{1+\hat z^2}{\hat z}
\right],
\end{align}
where $\hat \zeta^2 \equiv (2P_h \cdot \tilde v)^2/\tilde v^2$ (
$\hat \zeta \to \infty$ as $\tilde v^- \to 0$) regularizes
the rapidity divergence for the TMD FF.  

The contribution from $\Delta_F(z',z)$ (the third term in $[\ \ ]$ of (\ref{setupq}))
is obtained to be
\begin{align}\label{quark-dyn}
\left[ \Tr{} \Delta(z_f,\vec k_\perp) \frac{\Slash{w}}{2} 
\right]^{\mathrm{q-tw3}}_{\Delta_F}
 & = 
\frac{4\alpha_s M_h  \epsilon^{k_\perp P_h S_\perp w}}
{ (2\pi)^2P_h^- \vec k_\perp^4 z_f^2}
\left[
\int \frac{dz}{z} d\left(\frac{1}{z'}\right) \frac{1/z}{1/z-1/z'} 
\right. {\nonumber}\\
&\left.
\times \left\{
\Im \widehat D_{FT}(z,z') \left(
\frac{1}{1/z'}\frac{1}{2N}\frac{2-\hat z}{\hat z}
+\frac{1}{1/z'-1/z_f}\frac{N}{2}(1+\hat z)
\right)
\right. \right.{\nonumber}\\
&\left.
+\Im \widehat G_{FT} (z,z')\left(
\frac{1}{1/z'}\frac{1}{2N}-\frac{1}{1/z'-1/z_f}\frac{N}{2}(1-\hat z)
\right)
\right\} {\nonumber}\\
&\left.
-\int \frac{dz}{z}d\left(\frac{1}{z'}\right)\frac{C_F}{1/z-1/z'}
\left(\Im \widehat D_{FT}(z,z')- \Im \widehat G_{FT}(z,z') \right)(1+\hat z) 
\right. {\nonumber}\\
&\left.
- \int \frac{dz}{z}\frac{-2}{z} d\left(\frac{1}{z'}\right)
\frac{1}{(1/z-1/z')^2} \Im \widehat D_{FT}(z,z')\frac{N}{2}\frac{1+\hat z^2}{\hat z}
\right].
\end{align}

Taking the sum of
(\ref{quark-int}), \ref{quark-kin}) and (\ref{quark-dyn}) and
using the
EOM relation (\ref{EOMq}) and the LIR (\ref{LIRq}),
one obtains the twist-3 quark FF contribution to the polarizing FF
$D_{1T}^\perp$ as
\begin{align}\label{D1Tquark}
   D_{1T}^{\perp }(z_f,P_{T}^2) \Bigr|^{\mathrm{quark}}
=\frac{\alpha_s}{2\pi^2}
\frac{2M_h^2 z_f^2}{P_{T}^4} 
\left[ \int \frac{dz}{z} A 
+C_F D_{1T}^{\perp(1)}(z_f) 
\left(
\ln \frac{\hat \zeta^2}{P_{T}^2} 
-1 \right)
\right],
\end{align}
where 
$A$ is defined in (\ref{A}), which appeared in the 
small $q_T$-limit of the twist-3 quark FF contribution to the cross section for $ep\to e\Lambda^\uparrow X$.  

\subsubsection{Gluon fragmentation channel}

\begin{figure}[htb]
\begin{center}
\includegraphics[scale=1]{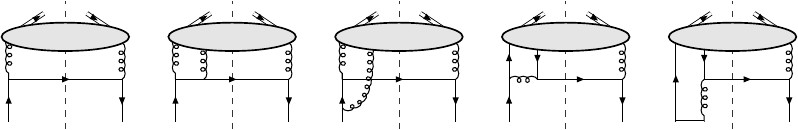}
\end{center}
\caption{
Feynman diagrams for the twist-3 gluon FF contribution to the
polarizing FF at large $P_T$.  Diagrams in which a gluon line coming out of the upper blob
directly couples to the Wilson line become either zero or suppressed 
at large $\hat \zeta^2$, so that they are not
shown here.  }
\label{diaggluonfrag}
\end{figure}

Next we proceed to calculate the twist-3 gluon FF contribution
to the TMD polarizing FF (\ref{setupg}).  
Relevant Feynman diagrams are shown in 
Fig. \ref{diaggluonfrag}.  

The contribution from $\widehat \Gamma (z)$ to (\ref{setupg}) (the first term)
can be obtained as
\begin{align}\label{naiveintg}
\left[ \Tr{} \Delta(z_f,\vec k_\perp) \frac{\Slash{w}}{2} 
\right]^{\mathrm{g-tw3}}_{\widehat\Gamma}
=\frac{2\alpha_s M_h C_F \epsilon^{k_\perp P_h S_\perp w}}
{(2\pi)^2P_h^- \vec k_\perp^4 z_f^2}
\int \frac {dz}{z}\frac{1}{z}\Delta \widehat G_{3\bar T}(z)
\frac{1-\hat z}{\hat z^2}2 (2-\hat z)\hat z^2.
\end{align}

The contribution from $\widehat \Gamma_\partial (z)$ to (\ref{setupg}) (the second term)
reads 
\begin{align}\label{naiveking}
\left[ \Tr{} \Delta(z_f,\vec k_\perp) \frac{\Slash{w}}{2} 
\right]^{\mathrm{g-tw3}}_{\widehat\Gamma_\partial}
& =
\frac{2\alpha_s M_h C_F \epsilon^{k_\perp P_h S_\perp w}}
{(2\pi)^2P_h^- \vec k_\perp^4 z_f^2}
\int \frac{dz}{z}z^2 \frac{1-\hat z}{\hat z^2}\left[
 (4-3\hat z+ \hat z^2) \widehat G_T^{(1)}(z) + (2-\hat z)\Delta \widehat H_{T}^{(1)}(z)
\right. {\nonumber}\\
& \left. 
- \left(
\frac{1-\hat z}{z} \frac{\partial \Delta \widehat H_T^{(1)}(z)}{\partial (1/z)}
+ \frac{2- 2 \hat z + \hat z^2}{z} \frac{\partial \widehat G_T^{(1)}(z)}{\partial (1/z)}
\right)
\right].
\end{align}

The contribution from $\widehat \Gamma_{FS}$ to (\ref{setupg}) 
reads
\begin{align}\label{naiveO}
\left[ \Tr{} \Delta(z_f,\vec k_\perp) \frac{\Slash{w}}{2} 
\right]^{\mathrm{g-tw3}}_{\widehat \Gamma_{FS}}
&=- \frac{ 4 \alpha_s M_h C_F
\epsilon^{k_\perp P_h S_\perp w} }{4 (2\pi)^2\vec k_\perp^4 z_f^2 P_h^-}
\int \frac{dz}{z}d\left(\frac{1}{z'}\right)
\frac{z'}{(1/z-1/z')^2} \frac{(1-\hat z)}
{ \hat z^2\hat z'}
{\nonumber}\\
&\times \Im \left[
 (-2 +\hat z) (\hat z^2 + 2\hat z'^2 - 2\hat z(1+\hat z'))
\widehat O_1 \left(\frac{1}{z'},\frac{1}{z}\right)
\right. {\nonumber}\\
&\left. 
+(2\hat z^3 - 2\hat z'^2 - 2\hat z^2(2+\hat z') + \hat z(4+2\hat z' + \hat z'^2))
\widehat O_2 \left(\frac{1}{z'},\frac{1}{z}\right)
\right. {\nonumber}\\
&\left.
+(-4\hat z^2 + \hat z^3 -2\hat z'^2 + \hat z(4+2\hat z' + \hat z'^2))
\widehat O_2 \left(\frac{1}{z}-\frac{1}{z'},\frac{1}{z}\right)
\right].
\end{align}
Using the second relation of (\ref{sym-property1})
and the fact that the change of the integration variables 
$1/z' \leftrightarrow 1/z-1/z'$ leaves $\int_0^{1/z} d \left(\frac{1}{z'}\right)$ invariant, 
together with the following relations
\begin{align}
&\frac{z'}{1/z-1/z'} = z'z + \frac{z}{1/z-1/z'}, &
 &\int d \left(\frac{1}{z'}\right) \frac{z'}{1/z-1/z'} \widehat O_1
= \int d \left( \frac{1}{z'}\right) \frac{2z}{1/z-1/z'}\widehat O_1,&
\label{methodOhat}
\end{align}
(\ref{naiveO}) can be rewritten as
\begin{align}\label{D1TgluonO}
\left[ \Tr{} \Delta(z_f,\vec k_\perp) \frac{\Slash{w}}{2} 
\right]^{\mathrm{g-tw3}}_{\widehat \Gamma_{FS}}
=- \frac{\alpha_s M_h \epsilon^{k_\perp P_h S_\perp w}}
{ (2\pi)^2 \vec k_\perp^4 P_h^- z_f^2} \int \frac{dz}{z}B_{\widehat O},
\end{align}
where $B_{\widehat O}$ represents the sum of the terms in (\ref{B})
containing $\Im \widehat O_{1,2}$, i.e., the same as the $\widehat O_{1,2}$ contribution
to the twist-3 cross section 
$\mathcal{F}_1|_{q_T\ll Q}$ in (\ref{F1lim}).

Likewise 
the contribution from $\widehat \Gamma_{FA}$ to (\ref{setupg}) 
can be obtained as 
\begin{align}
\left[ \Tr{} \Delta(z_f,\vec k_\perp) \frac{\Slash{w}}{2} 
\right]^{\mathrm{g-tw3}}_{\widehat \Gamma_{FA}}
& =
- \frac{4 \alpha_s M_h C_F\epsilon^{k_\perp P_h S_\perp w}}
{4(2\pi)^2 \vec k_\perp^4  z_f^2 P_h^-}
\int \frac{dz}{z} d\left(\frac{1}{z'}\right)
 \frac{1-\hat z}{\hat z^2}2z^2
{\nonumber}\\
& \times\Im\left[
\frac{1/z}{(1/z-1/z')^2} (- (-2+\hat z)^2)
\widehat N_1 \left(\frac{1}{z'},\frac{1}{z}\right)
\right. {\nonumber}\\
&\left.
+ \frac{1/z}{(1/z-1/z')^2} (-\hat z^2 + 4 \hat z-4)
\widehat N_2  \left(\frac{1}{z'},\frac{1}{z}\right)
\right.
{\nonumber}\\
&\left.
+\frac{1/z}{(1/z-1/z')^2} (2 \hat z^2 - 4 \hat z+4)
\widehat N_2 \left(\frac{1}{z}-\frac{1}{z'}, \frac{1}{z}\right)
\right], 
\label{Ncont}
\end{align}
where we have used the first relation of (\ref{sym-property1}).

The contribution from $\widetilde \Delta$
to (\ref{setupg}) (the last term) can be obtained as
\begin{align} 
\left[ \Tr{} \Delta(z_f,\vec k_\perp) \frac{\Slash{w}}{2} 
\right]^{\mathrm{g-tw3}}_{\widetilde \Delta}
& =
 - \frac{2\alpha_s M_h\epsilon^{k_\perp P_h S_\perp w}}
{ (2\pi)^2\vec k_\perp^4 z_f^2 P_h^- N}
\int \frac{dz}{z}d\left(\frac{1}{z'}\right) 
{\nonumber}\\
&\times
 \left\{
\Im \widetilde G_{FT} \left(\frac{1}{z'},\frac{1}{z'}-\frac{1}{z}\right)\left[
\frac{z}{1/z-1/z'}+ \frac{z_f^2}{-1+\hat z'}
\right] \right. {\nonumber}\\
&\left.+ \Im \widetilde D_{FT} \left(\frac{1}{z'},\frac{1}{z'}-\frac{1}{z}\right)
 z^2 \left[
\frac{1}{z}\frac{2-\hat z}{1/z-1/z'}\frac{-1}{\hat z}
+ \frac{-\hat z(-2+\hat z)}{1-\hat z'}
\right]
\right\}.
\label{DGtilde}
\end{align}

The total contribution from the gluon fragmentation channel to $D_{1T}^\perp$ 
can be obtained from the sum of
(\ref{naiveintg}), (\ref{naiveking}), (\ref{D1TgluonO}), (\ref{Ncont}) and 
(\ref{DGtilde}).   In taking the sum, one should note that
the intrinsic, kinematical, and dynamical twist-3 gluon FFs are related
among themselves by
the EOM relation (\ref{EOMg1}) and the LIRs, (\ref{EOMLIRg1}) and (\ref{EOMLIRg2}), 
except that $\Im\widehat O_{1,2}$ and $\Im\widetilde G_{FT}$
are independent from others.  
Here we employ the following procedure:
We express the intrinsic FF $\Delta \widehat G_{3\bar{T}}$ in terms of the others
by using (\ref{EOMg1}), and also eliminate the derivatives of the kinematical
FFs by using (\ref{EOMLIRg1}) and (\ref{EOMLIRg2}).  
This gives the final form for the twist-3 gluon FF contribution to $D_{1T}^\perp$ 
as
\begin{align}\label{polarizingTMDgluon}
  \left. D_{1T}^{\perp }(z_f,P_{T}^2)\right|^{\rm gluon}=\frac{\alpha_s}{2\pi^2}
\frac{2M_h^2 z_f^2}{P_{T}^4} 
\int \frac{dz}{z} \left(- \frac{1}{4}B \right), 
\end{align}
where $B$ is defined in (\ref{B}) which appeared 
in the $q_T\ll Q$-limit of the twist-3 cross section for $ep\to e\Lambda^\uparrow X$.

\subsection{TMD factorization formula for $ep\to e \Lambda^\uparrow X$
at $\Lambda_{\rm QCD}\ll P_T \ll Q$ }

Taking the sum of the twist-3 quark and gluon FF contributions, 
(\ref{D1Tquark}) and (\ref{polarizingTMDgluon}), 
the polarizing FF $D_{1T}^\perp (z_f, P_{T}^2)$ at $P_{T}\gg \Lambda_{QCD}$
can be written in terms of the twist-3 FFs as
\begin{align}\label{polarizingTMD}
  D_{1T}^{\perp }(z_f,P_{T}^2)=\frac{\alpha_s}{2\pi^2}
\frac{2M_h^2 z_f^2}{P_{T}^4} 
\left[ \int \frac{dz}{z} \left( A - \frac{1}{4}B \right)
+C_F D_{1T}^{\perp(1)}(z_f) 
\left(
\ln \frac{\hat \zeta^2}{P_{T}^2} 
-1 \right)
\right], 
\end{align}
where $A$ and $B$ are identical to those in (\ref{A}) and (\ref{B}).  
Inserting (\ref{softinverse}), (\ref{unpolTMD}) and (\ref{polarizingTMD})
into the TMD cross section formula
(\ref{TMDcrosssection}), 
one finally obtains
\begin{align}\label{finalresult}
 \frac{d^6 \sigma}{d x_{bj} dQ^2 dz_f dq_T^2 d \phi d\chi} 
&=
\frac{-4 \alpha_s M_h \sigma_0}{(2\pi)^2 q_T^3} \sin \Phi_S\sum_qe_q^2
\left[
q (x_{bj}) \int \frac{dz}{z} \left( A - \frac{1}{4}B\right) \right. \nonumber\\[5pt]
& \left. + C_F D_{1T}^{\perp (1)} (z_f)
\int \frac{dx}{x}q(x)
\frac{1+ \hat x^2}{(1- \hat x)_+}  
+ \frac{1}{2}D_{1T}^{\perp (1)}(z_f) \int \frac{dx}{x}G(x)(1-2\hat x + 2\hat x^2)
\right. {\nonumber}\\[7pt]
&\left.
+ C_F q (x_{bj}) D_{1T}^{\perp (1)}(z_f) 
\left\{
\left(\ln\frac{x_{bj}^2 \zeta^2}{q_T^2}-1\right)
+ \left(\ln \frac{\hat \zeta^2}{z_f^2 q_T^2}-1 \right)
- \left(\ln \rho^2 -2\right)
\right\}
\right].  
\end{align}
where
three terms in $\{\}$ come from
$q(x_{bj},q_T^2)$, $D_{1T}^\perp(z_f,z_f^2 q_T^2)$ and $S^{-1}(q_T^2)$.  
If one sets $x_{bj}^2 \zeta^2 = \hat \zeta^2 /z_f^2 = \rho Q^2$
as a regularization scheme for the rapidity divergence
\cite{Ji:2004wu,Ji:2006br}, the terms in $\{\}$ of
(\ref{finalresult})
become $2\ln \frac{Q^2}{q_T^2}$, and 
(\ref{finalresult}) completely agrees with
(\ref{F1}).  
This way we have confirmed that
the twist-3 FF contribution and the TMD polarizing FF contribution to
$ep\to e \Lambda^\uparrow X$ are consistent at 
$\Lambda_{\rm QCD} \ll P_{T}\ll Q$.  

\section{Conclusion}
In this paper we have investigated the consistency between the 
collinear twist-3 FF contribution and the TMD polarizing FF contribution
at $\Lambda_{\rm QCD}\ll P_T \ll Q$ 
for $ep\to e\Lambda^\uparrow X$ in LO QCD.  
Although the collinear twist-3 FFs give rise to all five 
azimuthal asymmetries for this process, only $\sin\Phi_S$-asymmetry
survives in the leading inverse power of $P_T/Q$.  
In the TMD factorization, the $\sin\Phi_S$-asymmetry is caused by the TMD polarizing FF.  
Calculating the dependence of 
the polarizing FF on the intrinsic transverse momentum $k_\perp$
in perturbative QCD at $k_\perp \gg \Lambda_{\rm QCD}$, 
the FF can be expressed in terms of
the collinear twist-3 FFs including purely gluonic ones.  
Inserting this expression into the TMD factorization
formula, we found the result completely agrees with that obtained
as the small $P_T$-limit of the collinear twist-3 FF contribution to the cross section.  
This indicates that the two frameworks describe the polarization
consistently in the whole kinematics region.

\section*{Acknowledgments}
This work has been supported by 
the establishment of Niigata university fellowships towards the creation of science
technology innovation (R.I.),
the Grant-in-Aid for
Scientific Research from the Japanese Society of Promotion of Science
under Contract Nos.~19K03843 and 24K07044 (Y.K.).  



\end{document}